\documentclass[aps,prl,showpacs,twoside,twocolumn,10pt,nofootinbib]{revtex4-2}
\usepackage[colorlinks=true, citecolor=blue, urlcolor=blue ]{hyperref}
\usepackage{epsfig,newlfont,amssymb,amsfonts,amsmath,bm,subfigure,palatino,mathtools,amsthm,braket, soul,enumitem,xcolor, graphics,graphicx,times,physics}
\usepackage[dvipsnames]{xcolor}
\usepackage{algorithm}
\usepackage{algpseudocode}
\usepackage{tikz}
\usetikzlibrary{quantikz}
\usepackage[normalem]{ulem}
\newtheorem{theorem}{Theorem}
\newtheorem{corollary}{Corollary}

\newtheorem{proposition}{Proposition}

\newtheorem{lemma}{Lemma}

\let\braket\relax
\newcommand{\braket}[1]{\left\langle #1 \right\rangle}

\begin{document}
\let\ORIGaddcontentsline\addcontentsline
\renewcommand{\addcontentsline}[3]{} 
\title{Fermionic non-Gaussianity via computable measures and experimentally accessible bounds}
\author{Leela Ganesh Chandra Lakkaraju}
\author{Naga Dileep Varikuti}
\author{Philipp Hauke}
\affiliation{Pitaevskii BEC Center, CNR-INO and Dipartimento di Fisica, Universit\`a di Trento, Via Sommarive 14, Trento, I-38123, Italy}	
\affiliation{INFN-TIFPA, Trento Institute for Fundamental Physics and Applications, Via Sommarive 14, Trento, I-38123, Italy}

\begin{abstract}
Interacting fermions owe both their non-trivial correlations and their computational power to their departure from Gaussianity. Yet, quantifying that departure geometrically faces two challenges: the Gaussian manifold is non-convex, and perturbative treatments are confined to weak interactions.  Here, we introduce a geometric measure of fermionic non-Gaussianity, the fidelity distance of a target state to the manifold of fermionic Gaussian states.  We construct an iteration formalism to numerically access it, applicable also at strong coupling.  Further, we prove a sharpened operator Jensen inequality, of interest in its own right. We use it to derive a witness that involves only two- and four-point fermionic correlators. Numerical benchmarks on the one-dimensional Fermi--Hubbard model show that the quality of the witness improves with system size and, for strong coupling, is tighter than antiflatness. Our results open a scalable route to charting non-Gaussianity of strongly-interacting fermionic quantum simulators.
\end{abstract}
\maketitle

\emph{Introduction.---}Strongly correlated quantum matter draws its richness from higher-order correlations that no effective single-particle picture can capture. Prominent examples are doped Fermi--Hubbard chains, whose antiferromagnetic order is recovered by string correlators~\cite{Alcaraz1992,Hilker2017}; quantum spin liquids, whose topological order resides in non-local quantum fluctuations~\cite{Savary2016,Hauke_spinliquids_2013,Wen_book,spin_liquid_review,Semeghni_experimental_2021,Satzinger2021,quantum_spin_liquids_2020}; and non-Fermi liquids, which cannot be described by simple quasiparticles~\cite{Stweart2001,StewartAddendum,Lee2018Mar_recent_review_non_fermi,Hartnoll,Abbamonte2022}. Fermionic Gaussian states occupy the opposite extreme. As ground states of quadratic Hamiltonians, they obey Wick's theorem, so that  the complexity of an exponentially large Hilbert space collapses to that of a covariance matrix of polynomial size~\cite{Bravyi2005-ue, Surace_Tagliacozzo_notes}; this renders them analytically tractable and efficiently simulable on a classical computer~\cite{Valiant2001,10.1098/rspa.2008.0189, divincenzo_terhal_2002, Jozsa_jordan_wigner, knill2001fermioniclinearopticsmatchgates}. Yet, this very transparency is a limitation: salient physics of interacting fermions and, indeed, their computational power reside in the departure from Gaussianity. Supplementing matchgate circuits with any pure non-Gaussian state promotes them to universal quantum computation~\cite{Josza_krauss_prl_fermionic_non_gaussian_2019}.  Various quantifiers for fermionic non-Gaussianity have been discussed~\cite{3yx4-1j27, haug2026practicaltestswitnessesfermionic, ares2026asymmetrylowerboundfermionic, coffman2025measuringnongaussianmagicfermions,tarabunga2026computablefermionicnongaussianitycovariance,poetri_prx,hoshino2026universalboundphasetransition}, but geometrically quantifying the departure of a quantum many-body system from the Gaussian states remains an outstanding challenge \cite{Bittel_Eisert_prx_quantum, Eisert_Hackl_local_optimisation}. Distance measures for non-Gaussianity are not expectation values of a fixed observable, but are defined through an optimization over the entire Gaussian manifold, which---in contrast to many other quantum resources~\cite{QRT_chitambar}---forms a non-convex set. It thus remains a pressing issue to identify measures that are at once operationally meaningful, geometrically faithful, and accessible to experiment~\cite{Turner2017, Pachos2022quantifying, 3yx4-1j27, coffman2025measuringnongaussianmagicfermions, varikuti2025deep}.

\begin{figure}
    \centering
    \includegraphics[width=0.9\linewidth]{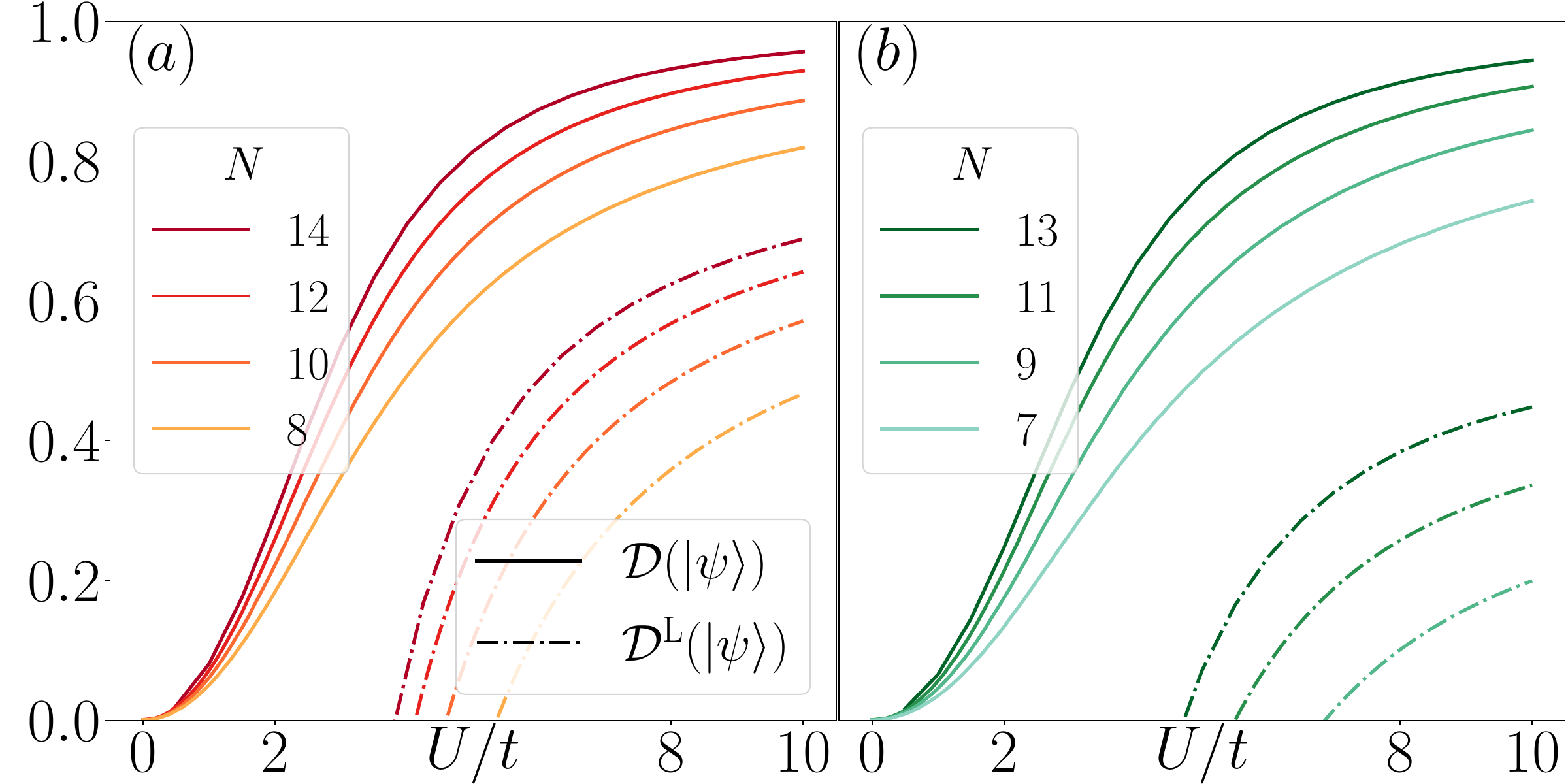}
        \caption{Non-Gaussianity of the ground state of the Fermi--Hubbard chain (solid curves), quantified by the distance to the closest Gaussian state ($\mathcal{D}(|\psi\rangle)$) obtained via the iterative procedure of Theorem~\ref{theorem1}. The $N$ fermionic modes are at [for even $N$, panel (a)] or near half-filling [odd $N$, (b)].
        For all $N$ considered, the non-Gaussianity increases with $U/t$. The lower bound obtained in Theorem~\ref{thm:lower-bound} (dashdotted) provides a genuine witness of non-Gaussianity at intermediate and strong interactions and becomes tighter with increasing $N$.}
    \label{fig:fidelity_min_lower_bound}
\end{figure}
In this Letter, we address this question by developing a practical approach to a geometric measure of fermionic non-Gaussianity: the fidelity distance from a given state to the closest fermionic Gaussian state. The Gaussian state that realizes this distance is of physical interest in its own right, for it constitutes the best free-fermion description of the target state~\cite{Turner2017}.  Locating this closest Gaussian state requires, however, maximizing the overlap over the Gaussian manifold. Our first result is a practical algorithm that implements this maximization. It is based on a self-consistent iteration that takes the fermionic two-point functions as input and generates Gaussian states that in our benchmarks  consistently have increasing overlap with the target. Second, we render the measure accessible to experiment. To this end, we prove a sharpened operator Jensen inequality \cite{jensen_original, HANSEN_PEDERSEN_2003, liao2019sharpening}, which may find relevance beyond the present context, and derive from it bounds that involve only two- and four-point correlators, quantities routinely measured in quantum-gas microscopes and on digital quantum processors \cite{google_fermion_2015,Bergschneider2019,alam2025programmabledigitalquantumsimulation}. We benchmark measure and bounds on non-perturbative ground states of the one-dimensional Fermi--Hubbard model, where non-Gaussianity builds up as the interaction is turned on (see Fig.~\ref{fig:fidelity_min_lower_bound}).

Our approach makes progress towards filling a hole within the resource theory of fermionic non-Gaussianity---enabling us to quantify the resource content of a state by its minimal infidelity with respect to the free states, a standard procedure for other quantum resource theories  \cite{QRT_chitambar}.  Besides this conceptual interest, the developed witness also has some practical advantages. Various previous approaches to witness non-Gaussianity are based on expansions, e.g., in two-mode entanglement energies~\cite{Pachos2022quantifying} or in correlations up to four-point~\cite{Froland2025}.  Our witness, in contrast, requires no expansion in the interaction strength, permitting to evaluate it in non-perturbative, strongly-correlated regimes.  We also benchmark our witness against antiflatness~\cite{3yx4-1j27}, which for pure states gives a lower bound on the distance to Gaussian states~\cite{haug2026practicaltestswitnessesfermionic}. We find that our witness performs less well at weak interactions, but is tighter at intermediate to strong coupling and---unlike antiflatness---improves with system size.  Our construction therefore presents a scalable approach to quantify non-Gaussianity, theoretically and experimentally, in challenging strongly-interacting regimes. 

\emph{Fidelity-based measure for non-Gaussianity.---}We define the distance-based measure for
non-Gaussianity~\cite{Turner2017, Eisert_Hackl_local_optimisation, Bittel_Eisert_prx_quantum,Dias2024classicalsimulation} as
\begin{equation}
    \mathcal{D}(\rho) = \min_{\sigma \in \mathcal{G}} \left[ 1-\mathcal{F}(\rho,\sigma) \right],
    \label{eq:witness_non_gaussian}
\end{equation}
where $\mathcal{G}$ denotes the non-convex set of fermionic Gaussian states and
$\mathcal{F}(\rho,\sigma) = \left( \text{Tr}\sqrt{\sqrt{\rho}\,\sigma\sqrt{\rho}} \right)^{2}$
is the Uhlmann--Jozsa fidelity~\cite{UHLMANN1976,Jozsa_fidelity}.\footnote{A computationally lighter alternative is the
Hilbert--Schmidt norm based fidelity $\mathcal{F}_{2}(\rho,\sigma) = \text{Tr}(\rho\sigma)/
\max\!\left(\text{Tr}\rho^{2},\text{Tr}\sigma^{2}\right)$~\cite{Liang_fidelity_review}. }
The fidelity distance carries operational meaning. Calling $\sigma^{\star}$ the Gaussian state closest to $\rho$, no measurement can distinguish it from the target with success probability exceeding $\tfrac12\bigl[1+\sqrt{\mathcal{D}(\rho)}\bigr]$
~\cite{Helstrom1969,FuchsvandeGraaf1999}. The distance also bounds the error incurred on any bounded observable, $|\langle O\rangle_\rho-\langle O\rangle_{\sigma^{\star}}|\le2\lVert O\rVert_\infty\sqrt{\mathcal{D}(\rho)}$ ~\cite{NielsenChuang}, so a small distance certifies that the free-fermion description approaches the correlations of the target at every order, not just the quadratic ones. The measure is, moreover, stable: $\sqrt{\mathcal{D}}$ is the distance from the target to $\mathcal{G}$ in the purified metric~\cite{Tomamichel2010}, so a state prepared to within $\epsilon$ of the intended one in that metric carries a distance shifted by at most $\epsilon$.
\begin{proposition}
\label{prop1}
$\mathcal{D}(\rho)$ is a faithful monotone of the resource theory of fermionic
non-Gaussianity.
\end{proposition}
\begin{proof}
    The proposition rests on two properties of the distance derived from the Uhlmann fidelity. The first is faithfulness: $\mathcal{F}(\rho,\sigma) = 1$ iff $\rho = \sigma$, so that $\mathcal{D}(\rho) = 0$ exactly on $\mathcal{G}$. Any $\mathcal{D}(\rho) > 0$ witnesses the non-Gaussianity of $\rho$. The second is monotonicity, $\mathcal{D}(\mathcal{E}(\rho)) \le \mathcal{D}(\rho)$, where $\mathcal{E}$ is a Gaussian operation, i.e., a completely positive trace-preserving map that takes every Gaussian state to a Gaussian state, $\mathcal{E}(\mathcal{G}) \subseteq \mathcal{G}$; monotonicity follows from the data-processing inequality $\mathcal{F}(\mathcal{E}(\rho), \mathcal{E}(\sigma)) \ge \mathcal{F}(\rho,\sigma)$ together with $\mathcal{E}(\sigma) \in \mathcal{G}$ (see SM \cite{SM}). 
\end{proof}

The measure is therefore well-defined and carries resource-theoretic meaning for arbitrary mixed states. Hereafter, we restrict the discussion to pure non-Gaussian states $\rho = |\psi\rangle\langle\psi|$, which constitute the essential resource for universal quantum computation within the matchgate paradigm~\cite{Josza_krauss_prl_fermionic_non_gaussian_2019}. For pure targets, the fidelity reduces to $\mathcal{F}(\rho,\sigma) = \langle\psi|\sigma|\psi\rangle$.

Every fermionic Gaussian state can be expressed as the Gibbs state of a Hamiltonian that is quadratic in the Majorana operators $\gamma_{2j-1}=c_j+c^\dagger_j$ and $\gamma_{2j}=-i(c_j-c^\dagger_j)$, which obey $\{\gamma_a,\gamma_b\}=2\delta_{ab}$. We therefore write
\begin{equation} 
    \sigma_K=\frac{e^{-\Gamma_K}}{\mathcal{Z}_K},
    \quad
    \Gamma_K=\frac{i}{4}\sum_{a,b}K_{ab}\,\gamma_a\gamma_b
            =\frac{1}{2}\sum_{a<b}K_{ab}\,X_{ab},
    \label{eq:gaussian-state}
\end{equation}
with $K$ real and antisymmetric, $\mathcal{Z}_K=\operatorname{Tr}e^{-\Gamma_K}$, and $X_{ab}=i\gamma_a\gamma_b$. The $X_{ab}$ are Hermitian and square to unity. An orthogonal rotation of the Majorana operators brings $K$ to $2\times2$ blocks and diagonalizes $\Gamma_K$, $\Gamma_K=\sum_k\beta_k\Big(n_k-\frac{1}{2}\Big)$. Here, $n_k$ are the occupations of the associated Bogoliubov modes and $\beta_k=\ln[(1-m_k)/(1+m_k)]$, with $m_k$ the eigenvalues of the covariance matrix $M_{ab}=\operatorname{Tr}(\sigma_KX_{ab})$~\cite{Surace_Tagliacozzo_notes}. 

Finding the closest Gaussian state means maximizing $\mathcal{F}(K)=\bra{\psi}\sigma_K\ket{\psi}$ over $K$. We now show that this reduces to a self-consistency condition on the covariance matrix.

\begin{lemma}
\label{thm:stationarity}
Let $\ket{\psi}$ be a pure target state and let $X_{ab}=i\gamma_a\gamma_b$. Stationarity of the fidelity $\mathcal{F}(K)$ with respect to the independent parameters $K_{ab}$ $(a<b)$ is equivalent to the self-consistency condition
\begin{equation}
    M_{ab}=
    \displaystyle\int_{0}^{1}\mathrm{d}s
     \frac{ \bra{\psi}e^{-s\Gamma}\,X_{ab}e^{-(1-s)\Gamma}\,\ket{\psi} }
    {\bra{\psi}e^{-\Gamma}\ket{\psi}}. 
    \label{eq:stationarity}
\end{equation}
\end{lemma}

\begin{proof}
Define $A=\bra{\psi}e^{-\Gamma}\ket{\psi}$, so $\mathcal{F}=A/\mathcal{Z}_K$. 
From Eq.~\eqref{eq:gaussian-state}, 
$\partial\Gamma/\partial K_{ab}=X_{ab}/2$, so Duhamel's formula~\cite{Wilcox1967Apr} gives
\begin{equation}
    \frac{\partial e^{-\Gamma}}{\partial K_{ab}}
    =-\frac{1}{2}\int_0^1\!\mathrm{d}s\;
      e^{-s\Gamma}\,X_{ab}\,e^{-(1-s)\Gamma}.
    \label{eq:duhamel}
\end{equation}
Under a trace, the two exponentials recombine by cyclicity, yielding 
$\partial\mathcal{Z}_K/\partial K_{ab}
 =-\tfrac12\operatorname{Tr}(X_{ab}e^{-\Gamma})
 =-\tfrac12\mathcal{Z}_KM_{ab}$. Inside $\bra{\psi}\cdots\ket{\psi}$ no such
recombination is possible unless $[\Gamma,X_{ab}]=0$, and generically the integral
survives. Defining 
$g_{ab}(s)=\bra{\psi}e^{-s\Gamma}X_{ab}e^{-(1-s)\Gamma}\ket{\psi}$ and
$N_{ab}=\int_0^1\mathrm{d}s\,g_{ab}(s)$, one has
$\partial A/\partial K_{ab}=-\tfrac12N_{ab}$. The quotient rule gives
\begin{equation}
    \frac{\partial\mathcal{F}}{\partial K_{ab}}
    =\frac{1}{\mathcal{Z}_K}\frac{\partial A}{\partial K_{ab}}
     -\frac{A}{\mathcal{Z}_K^{2}}\frac{\partial\mathcal{Z}_K}{\partial K_{ab}}
    =\frac{\mathcal{F}}{2}\left(M_{ab}-\frac{N_{ab}}{A}\right).
    \label{eq:gradient}
\end{equation}
Since $\sigma_K$ is positive definite, $\mathcal{F}\geq 0$, and the non-trivial stationary points are the solutions of $M_{ab}=N_{ab}/A$, which is Eq.~\eqref{eq:stationarity}. 
\end{proof}
% Equation~\eqref{eq:stationarity} is a self-consistency condition. Write $\ket{\psi_s}=e^{-s\Gamma}\ket{\psi}$ for the target state propagated in imaginary time by the trial Hamiltonian. The right-hand side of Eq.~\eqref{eq:stationarity} is then $\int_0^1\mathrm{d}s\,\bra{\psi_s}X_{ab}\ket{\psi_{1-s}}
% /\langle \psi_{1/2}|\psi_{1/2} \rangle$, the correlations of the target after
% propagation generated by the trial state. Stationarity demands that these reproduce the covariance matrix of that same trial state, i.e., the effective single-particle Hamiltonian and the correlations it generates must agree~\cite{RingSchuck1980}. 

Equation~\eqref{eq:stationarity} is a self-consistency condition, whose solutions encode the local and global minima of $\mathcal{D}$. Write $\ket{\psi_s}=e^{-s\Gamma}\ket{\psi}$ for the target propagated in imaginary time by the trial Hamiltonian. The right-hand side of Eq.~\eqref{eq:stationarity} is then the ordered average $\int_0^1\mathrm{d}s\,\braket{\psi_s|X_{ab}|\psi_{1-s}} /\braket{\psi_{1/2}|\psi_{1/2}}$, which is real because the integrands at $s$ and $1-s$ are complex conjugates. Stationarity demands that this average returns the covariance matrix of the very Gaussian state that generated the propagation. 

\begin{theorem}
\label{theorem1}
Evaluating the integral in Eq.~\eqref{eq:stationarity} by the trapezoidal rule yields the update, from iteration $r\to r+1$, 
\begin{equation}
    M^{(r+1)}_{ab}
    \;=\;
    \operatorname{Re}\!\left[
      {\bra{\psi}\,X_{ab}\,e^{-\Gamma^{(r)}}\,\ket{\psi}}/
           {\bra{\psi}\,e^{-\Gamma^{(r)}}\,\ket{\psi}}
    \right].
    \label{eq:update}
\end{equation}
\end{theorem}

\begin{proof}
For details, see SM \cite{SM}.
Hermiticity of
$\Gamma$ and $X_{ab}$ gives
$g_{ab}(s)^{*}=\bra{\psi}e^{-(1-s)\Gamma}X_{ab}e^{-s\Gamma}\ket{\psi}
 =g_{ab}(1-s)$. Thus, approximating $N_{ab}$ by the trapezoidal rule gives %the two endpoints of $g_{ab}$ are complex conjugates of one another,
\begin{equation}
    \tfrac{1}{2}\big[g_{ab}(0)+g_{ab}(1)\big]
    =\operatorname{Re}g_{ab}(0)
    =\operatorname{Re}\bra{\psi}X_{ab}e^{-\Gamma}\ket{\psi}.
    \label{eq:endpoints}
\end{equation}
Using this in Eq.~\eqref{eq:stationarity}, the update rule follows. 
\end{proof}

The trapezoidal rule evaluates the integral in Eq.~\eqref{eq:stationarity} only at its endpoints, yet it leaves the solutions untouched. The integrand in Eq.~\eqref{eq:stationarity} contains a Majorana bilinear built from the Bogoliubov modes of $\Gamma$. These merely acquire a factor $e^{-s\lambda}$ when moved through $e^{-s\Gamma}$, with $\lambda$ fixed by the two mode energies. Exact and trapezoidal gradients thus differ, component by component in this basis, by a positive factor between unity and $\beta_{\max}\coth\beta_{\max}$, where $\beta_{\max}=\max_k|\beta_k|$; in particular, for any finite $\beta_{\max}$ they vanish together. Consequently, Eqs.~\eqref{eq:stationarity} and~\eqref{eq:update} have identical solutions, i.e., the trapezoidal rule may alter the path of the iteration, but not its endpoint (see SM \cite{SM}).

Below, we seed the iteration with the Gaussian projection $\sigma^{(0)}$, the
Gaussian state whose covariance matrix matches that of the target,
$M^{(0)}_{ab}=\bra{\psi}X_{ab}\ket{\psi}$. 
In our benchmarks, few iterations suffice to converge. Two caveats are in order. The Gaussian manifold is not convex, and $\mathcal{F}$ is maximised on its boundary, where the pure Gaussian states saturate $|m_k|=1$. For numerical stability, we hold $|m_k|\le 1-\epsilon$ and check that the converged fidelity does not depend on $\epsilon$. Moreover, the iteration might converge to a local rather than a global optimum, in which case the obtained value overestimates $\mathcal{D}$. In benchmarks against simulated annealing \cite{original_simulated_annealing_kirkpatrick_1983} over the Gaussian manifold, we find the iteration to reach equal or better fidelities at a fraction of the cost (see SM \cite{SM}).

\emph{From distance measure to measurable bounds.}---Evaluating Eq.~\eqref{eq:update} requires the full many-body wavefunction. We now show that the distance can instead be bounded from below using only two- and four-point fermionic correlators of the target---quantities routinely measured in fermionic quantum simulators. The key tool is a sharpened operator Jensen inequality.

\begin{theorem}
\label{thm:jensen-main}
Let $A$ be a bounded self-adjoint operator with spectral edges
$\lambda_{\min/\max}$, and $\ket{\psi}$ a pure state with
$\mu_A=\bra{\psi}A\ket{\psi}$ and
$\operatorname{Var}_\psi(A)=\bra{\psi}A^{2}\ket{\psi}-\mu_A^{2}$. Let $f$ be
such that $h(x;\mu)=[f(x)-f(\mu)]/(x-\mu)^{2}-f'(\mu)/(x-\mu)$ is
non-decreasing in $x$ on $[\lambda_{\min},\lambda_{\max}]$, as holds for
$f=\exp$. Then
\begin{align}
    &h(\lambda_{\min};\mu_A)\operatorname{Var}_\psi(A) \le\;
    \bra{\psi}f(A)\ket{\psi}-f(\mu_A)\nonumber \\
    \;&\qquad\qquad\qquad\qquad\qquad
    \;\le\;
    h(\lambda_{\max};\mu_A)\operatorname{Var}_\psi(A).
    \label{eq:jensen}
\end{align}
\end{theorem}
\begin{proof}
See End Matter. Equation~\eqref{eq:jensen} sharpens the operator Jensen
inequality~\cite{HANSEN_PEDERSEN_2003} by resolving the Jensen gap to leading
order in the fluctuations, in the spirit of its scalar
counterpart~\cite{liao2019sharpening}.
\end{proof}

Applied to the fidelity, Eq.~\eqref{eq:jensen} yields the following bound.

\begin{theorem}
\label{thm:pointwise}
Let $\lambda_{\max}$ be the largest eigenvalue of $-\Gamma$, which is fixed by
$K$ alone, and let $\mu=-\bra{\psi}\Gamma\ket{\psi}$ and
$\operatorname{Var}_\psi(\Gamma)=\bra{\psi}\Gamma^{2}\ket{\psi}
-\bra{\psi}\Gamma\ket{\psi}^{2}$ be the first two moments of $\Gamma$ in the
target, which involve only its two- and four-point correlators. Then, for
every $K$,
\begin{equation}
    \mathcal{F}(K)\;\le\;B(K)\;\equiv\;
    \frac{e^{\mu}+h(\lambda_{\max};\mu)\operatorname{Var}_\psi(\Gamma)}
         {\mathcal{Z}_K}.
    \label{eq:bound}
\end{equation}
\end{theorem}
\begin{proof}
Apply the upper half of Eq.~\eqref{eq:jensen} with $A=-\Gamma$ and $f=\exp$,
for which $\mu_A=\mu$ and
$\operatorname{Var}_\psi(A)=\operatorname{Var}_\psi(\Gamma)$. This gives
$\bra{\psi}e^{-\Gamma}\ket{\psi}\le e^{\mu}
+h(\lambda_{\max};\mu)\operatorname{Var}_\psi(\Gamma)$; dividing by
$\mathcal{Z}_K>0$ yields Eq.~\eqref{eq:bound}.
\end{proof}

Equation~\eqref{eq:bound} holds at every $K$, yet optimising it over all of
them does not bound $\mathcal{D}$. The supremum defining $\mathcal{D}$ is
attained on a \emph{pure} Gaussian state~\cite{SM}, which carries $|m_k|=1$
and is reached only as $\beta_k\to\pm\infty$; there $B$ diverges, and the
unrestricted optimisation returns a trivial bound~\cite{SM}. A compact
restriction is unavoidable. Its price, however, can be stated exactly.

\begin{theorem}
\label{thm:lower-bound}
Let $\mathcal{B}_{\epsilon}=\{\sigma_K:\,|m_{k}(K)|\le1-\epsilon\ \,\forall k\}$ for
$\epsilon\in(0,1)$, with $N$ the number of fermionic modes. Then
\begin{equation}
    \mathcal{D}(\ket{\psi})\;\ge\;\mathcal{D}^{\mathrm{L}}_{\epsilon}
    \;\equiv\;1-\Big(1-\tfrac{\epsilon}{2}\Big)^{-N}
    \sup_{K\in\mathcal{B}_{\epsilon}}B(K).
    \label{eq:DL}
\end{equation}
\end{theorem}
\begin{proof}
Write $\mathcal{F}_{\mathcal{G}}
=\sup_{\sigma\in\mathcal{G}}\bra{\psi}\sigma\ket{\psi}
=1-\mathcal{D}(\ket{\psi})$, attained on a pure Gaussian state
$\ket{\phi}$~\cite{SM}. We take the single-particle basis in which the covariance matrix is block diagonal
and the state factorises as $\ket{\phi}=\bigotimes_{k}\ket{c_{k}}$. Build, in that same frame, $
    \sigma_{\epsilon}=\bigotimes_{k}\rho_{k}, $ $
    \rho_{k}=\Big(1-\tfrac{\epsilon}{2}\Big)\ket{c_{k}}\!\bra{c_{k}}
            +\tfrac{\epsilon}{2}\ket{\bar c_{k}}\!\bra{\bar c_{k}},
 $
with $\bar c_{k}$ the opposite occupation, so that $|m_{k}|=1-\epsilon$ and
$\sigma_{\epsilon}\in\mathcal{B}_{\epsilon}$. Discarding the second term of
each $\rho_{k}$ removes a positive semidefinite operator, whence
$\rho_{k}\ge(1-\tfrac{\epsilon}{2})\ket{c_{k}}\!\bra{c_{k}}\ge0$ and 
$\sigma_{\epsilon}\ge(1-\tfrac{\epsilon}{2})^{N}
\ket{\phi}\!\bra{\phi}$. Sandwiching with $\ket{\psi}$,
$    \bra{\psi}\sigma_{\epsilon}\ket{\psi}\;\ge\;
    \Big(1-\tfrac{\epsilon}{2}\Big)^{N}\mathcal{F}_{\mathcal{G}}\,:
$
retreating from the pure optimum into $\mathcal{B}_{\epsilon}$ costs a factor
$(1-\epsilon/2)^{N}$. As $\sigma_{\epsilon}$ is one element of $\mathcal{B}_{\epsilon}$, its overlap is at most the supremum over the whole set, and $\mathcal{F}_{\mathcal{G}}
\le(1-\tfrac{\epsilon}{2})^{-N}\sup_{K\in\mathcal{B}_{\epsilon}}
\mathcal{F}(K)$. Theorem~\ref{thm:pointwise} bounds $\mathcal{F}(K)$ by
$B(K)$ at every $K$.
\end{proof}

\begin{corollary}
\label{cor:eps}
Every $\epsilon$ yields a valid bound, and hence so does
$\mathcal{D}^{\mathrm{L}}\equiv\sup_{\epsilon\in(0,1)}
\mathcal{D}^{\mathrm{L}}_{\epsilon}$, which carries no free parameter and
satisfies $\mathcal{D}^{\mathrm{L}}\ge0$ for every state.
\end{corollary}
\begin{proof}
As $\epsilon\to1$, $\mathcal{B}_{\epsilon}$ contracts to $K=0$, where
$\sigma_{K}=\mathbb{I}/2^{N}$, $\mu=\operatorname{Var}_\psi(\Gamma)=0$ and
$h\to e^{\mu}/2$, so that $B=2^{-N}$ against a
prefactor $2^{N}$. The two cancel identically for every target,
$\mathcal{D}^{\mathrm{L}}_{\epsilon\to1}=0$, and the supremum is therefore
non-negative. It is attained at an interior $\epsilon^{\star}$ whenever it is
positive, where $\mathcal{D}^{\mathrm{L}}_{\epsilon}$ is stationary in
$\epsilon$. 
\end{proof}

Three features of this construction deserve emphasis. First, the regulator is not a technicality but the mechanism that makes the relaxation usable:  shrinking $\mathcal{B}_{\epsilon}$ suppresses $\sup B$ while the prefactor $(1-\epsilon/2)^{-N}$ grows, the two balance at $\epsilon^{\star}$, and the $\epsilon\to1$ endpoint pins $\mathcal{D}^{\mathrm{L}}$ at zero from below. In our benchmarks, we find the maximum over $\epsilon$ to be rather shallow; the bound at the fixed value $\epsilon=0.02$ already lies close to optimum $\epsilon^{\star}$~\cite{SM}; thus, we use $\epsilon=0.02$ and drop the $\epsilon$ subscript for brevity. The bound turns into a strictly positive witness of non-Gaussianity once the target is correlated enough for the variance term to dominate. Second, two- and four-point correlators appear, but Eq.~\eqref{eq:bound} does not descend from a Taylor expansion in the interaction strength; the bound is non-perturbative and remains informative where expansions lose control. Third, the relaxation deforms the optimisation landscape, so that the Gaussian state maximising $B$ need not coincide with the one maximising the exact fidelity. Locating it calls for its own stationarity condition, which we now give.

\begin{theorem}
\label{thm:bound-iteration}
Let $M^{\psi}_{ab}=\bra{\psi}X_{ab}\ket{\psi}$ be an element of the covariance
matrix of the target, $\bar{M}_{ab}$ the corresponding one of the eigenvector
of $-\Gamma$ at $\lambda_{\max}$, assumed non-degenerate, and
\begin{equation}
    S_{ab}=\tfrac{1}{2}\bra{\psi}\big\{-\Gamma,\,X_{ab}\big\}\ket{\psi}
    \label{eq:S-tensor}
\end{equation}
The matrix built from the four-point Majorana correlators of the target.
Writing $\alpha=e^{\mu}+h\operatorname{Var}_\psi(\Gamma)$ and evaluating $h$ and its partial derivatives at $(\lambda_{\max};\mu)$, the gradient of the bound reads
\begin{equation}
    \frac{\partial B}{\partial K_{ab}}
    =\frac{B}{2}\Big[\,M_{ab}
      -\big(c_0M^{\psi}+c_1\bar{M}+c_2S\big)_{ab}\Big],
    \label{eq:Bgradient}
\end{equation}
with the scalar coefficients
$
    c_0=(e^{\mu}-2\mu h
              +\operatorname{Var}_\psi(\Gamma)\,\partial_\mu h)/\alpha, c_1=(\operatorname{Var}_\psi(\Gamma)\,\partial_\lambda h)/\alpha,$ and  
    $c_2=2h/\alpha.$
At interior points of $\mathcal{B}_{\epsilon}$, $B$ is thus stationary if and
only if
\begin{equation}
    M=c_0\,M^{\psi}+c_1\,\bar{M}+c_2\,S,
    \label{eq:bound-selfconsistency}
\end{equation}
which, read as a map $M^{(r)}\mapsto M^{(r+1)}$, defines the update rule. The
proof is given in the SM~\cite{SM}.
\end{theorem}

Crucially, the entire iteration requires only $M^{\psi}$, $\mu$,
$\operatorname{Var}_\psi(\Gamma)$ and $S$---the two- and four-point Majorana
correlations of the target. These need to be measured only \emph{once}. Every
subsequent iteration re-contracts the same fixed data with an updated $K$, in
purely classical post-processing, so that locating the optimal reference state
demands no further access to the quantum tomography of the state.

\emph{The Fermi-Hubbard model.---}We demonstrate our main results for the Fermi--Hubbard (FH) model, the paradigmatic model for interacting fermions and a workhorse in cold-atom quantum simulation 
\cite{Zoller_fermi_hubbard_2019, optical_lattices_fermions_review, Dutta2015, Bluvstein_zoller_fermi_2023, 
Franz_nature, Xu2025Jun, Bojovic2026Apr}. Its Hamiltonian in one dimension reads 
\begin{equation}
    H_\mathrm{FH} = t\sum_{\langle i j \rangle,\sigma} c^\dagger_{i\sigma}c_{j\sigma}+c_{j\sigma}^\dagger c_{i\sigma} + U\sum_i  n_{i\uparrow}n_{i\downarrow}\,.
\end{equation}
Here, $c^\dagger_{i\sigma}$ ($c_{i\sigma}$) creates (annihilates) a fermion with spin $\sigma$ at site $i$, $t$ is the nearest-neighbor hopping amplitude, $n_{i\sigma}=c^\dagger_{i\sigma}c_{i\sigma}$, and $U$ denotes the on-site interaction strength. For $U=0$, the model reduces to a free-fermion system whose ground state is Gaussian.  By employing the above framework, we demonstrate how non-Gaussianity builds up as $U$ is increased. 

The corresponding numerical results for filling $\lfloor N/2 \rfloor$ are shown in Fig.~\ref{fig:fidelity_min_lower_bound}. For all system sizes considered, the geometric measure of non-Gaussianity $\mathcal{D}(|\psi\rangle)$ (solid lines) grows monotonically with interaction strength $U/t$. Moreover, for $U/t$ fixed, $\mathcal{D}(|\psi\rangle)$ in our data increases with system size. As discussed in the SM \cite{SM}, in these benchmarks our iterative algorithm consistently outperforms optimization via simulated annealing in both convergence speed and accuracy.  

In addition, Fig.~\ref{fig:fidelity_min_lower_bound} shows the lower bound obtained in Theorem~\ref{theorem1} (dashdot lines). As the bound relies on a mathematical relaxation via the operator inequality, it is not guaranteed to be strictly positive. We find it to become a non-trivial, positive witness of non-Gaussianity once the interaction strength surpasses a certain threshold. At the system sizes considered, the threshold is sensitive to the parity of the lattice, lying at lower $U/t$ for even-site configurations. More importantly, it moves towards smaller interactions as system size increases. Thus, away from the weakly-interacting regime, this experimentally viable bound can serve as a useful and scalable approximation to the exact geometric measure.

% \begin{figure}[H]
%     \includegraphics[width=\linewidth]{gauss_fidelity.pdf}
%     \caption{Upper and lower bounds of Fidelity when the Gaussian part of the state is considered. Here $L = 8$, $N_{\uparrow} = 4$ and $N_{\downarrow} = 4$. }
%     \label{fig:placeholder}
% \end{figure}

%\textit{Experimental relevance.---}
\emph{Experimental relevance.}---Theorem~\ref{thm:bound-iteration} permits measuring fermionic non-Gaussianity without any weak-coupling expansion, and at the same time trades full-state tomography for two- and four-point fermionic correlators, acquired only once and re-contracted classically at every subsequent iteration. Quantum gas microscopes measure such correlators in the Fermi--Hubbard model~\cite{Parsons2016,Boll2016,Cheuk2016} and have steadily resolved them to higher order~\cite{Koepsell2019,Koepsell2021,Franz_nature,franz_pnas}.  Generic correlators, required for $M^{\psi}$ and $S$, can be recovered from free, translationally invariant quenches followed by occupation-number measurements~\cite{Denzler2024}.

Fermi--Hubbard models are now also routinely realized on digital platforms at scales that start to reach beyond exact classical simulation, e.g., on superconducting qubits~\cite{Barends2015Jul_superconducting_fermihubbard, Stanisic2022Oct_superconducting_fermi_hubbard,pollman_supercond_fermions_2025,arute2020observationseparateddynamicscharge,alam2025programmabledigitalquantumsimulation,Chowdhury2026_fermihubbard_superconducting} and on trapped ions~\cite{alam2025fermionicdynamicstrappedionquantum}, where  off-diagonal pairing correlations %, inaccessible to local density measurements, 
have recently been measured directly~\cite{granet2026superconductingpairingcorrelationstrappedion}. 
Random matchgate shadows can supply the full set of correlations up to fourth order  at polynomial cost~\cite{Zhao2021,Low2024}, showing that non-Gaussianity can be certified efficiently using only Gaussian operations. 
Similar shadows have already been run on trapped-ion hardware within quantum--classical auxiliary-field quantum Monte Carlo calculations~\cite{Huang2024MatchgateQCQMC,Zhao2026QCAFQMC}, and the same snapshots give $\bra{\psi}\sigma_K\ket{\psi}$ for any Gaussian $\sigma_K$~\cite{Wan2023}. 

Semiconductor quantum-dot arrays~\cite{Hensgens2017,Dehollain2020} offer a small-scale testbed for accessing the optimised bound in Eq.~\eqref{eq:DL}. 
In such a setting, each quantity entering the derivative of $\mathcal{D}^{\mathrm{L}}$ via Eq.~\eqref{eq:gradient} comes with a rigorous bound on the number of snapshots required~\cite{Zhao2021,Denzler2024,Wan2023}, so their error bars carry over into an error bar on $\mathcal{D}^{\mathrm{L}}$.

\emph{Discussion.}---%Fermionic non-Gaussianity has largely been approached as a property to be detected. 
The geometry developed here makes fermionic non-Gaussianity a distance that can be measured: the departure of a state from the free-fermion manifold is bounded from both sides by quantities a quantum simulator can record, with no expansion in the interaction strength. That departure has been studied before. In particular, Ref.~\cite{Dias2024classicalsimulation} introduced the Gaussian fidelity and demonstrated several fundamental properties.\footnote{A related quantity is the interaction distance~\cite{Turner2017}, which considers states with free-particle spectra rather than Gaussian states on fixed modes, and which is bounded through the violation of Wick's theorem~\cite{Pachos2022quantifying}.} Here, we achieved three things. First, we showed that the non-Gaussianity distance is a faithful monotone of the resource theory. Second, we addressed a main difficulty of using fidelity-based measures of non-Gaussianity: a practical approach to the closest Gaussian state. 
Third, we introduced a witness that stays sharp once correlations are strong and the cumulant hierarchy cannot be truncated~\cite{Froland2025}. 

Benchmarked against fermionic antiflatness~\cite{3yx4-1j27}, which for pure states brackets the fidelity  distance~\cite{haug2026practicaltestswitnessesfermionic}, our witness $\mathcal{D}^{\mathrm{L}}$ is the looser bound at weak interactions, but takes over beyond intermediate coupling. The reason is structural: the bound due to the fermionic antiflatness contains a prefactor $1/(4N)$, which is essentially optimal~\cite{haug2026practicaltestswitnessesfermionic}; so that bound can never exceed $1/4$, however non-Gaussian the state. In contrast, $\mathcal{D}^{\mathrm{L}}$ carries no such ceiling; we find that it continues to track $\mathcal{D}$ and sharpens with system size. 

In the benchmark model, the non-Gaussianity witness sharpens with system size in precisely the strongly correlated regime. More broadly, the distance to the Gaussian manifold is what sets the cost of a free-fermion description. Thus, the value certified in an experiment indicates how far the created state has moved beyond the reach of classical matchgate calculations---a question that grows pressing, as quantum simulators are set to reach regimes where numerics can no longer serve as a benchmark. 
%Studying the thermalization of the non-Gaussianity, its interplay with quantum chaos~\cite{Varikuti2026impactofclifford}, and the emergence of other quantum resources required for quantum advantage is a compelling future direction~\cite{w97w-7zny}.

\acknowledgements
\textit{Acknowledgments.---}This project has been funded by the Caritro Foundation. 
This work was supported by the Swiss State Secretariat for Education, Research and Innovation (SERI) under contract number UeMO19-5.1. 
This work was supported by the Provincia Autonoma di Trento, and Q@TN, the joint lab between the University of Trento, FBK—Fondazione Bruno Kessler, INFN—National Institute for Nuclear Physics, and CNR—National Research Council, Italy. 

\textit{AI statement.---}%Besides for checking the manuscript and for improving writing, generative AI (Claude Fable 5.0, Opus 5.5) was used to prove the trapezoidal rule yields correct stationary states and to derive the regularization $(1-\epsilon/2)^{-N}$ in the witness, which avoids potential overestimation.
Generative AI (Claude Fable 5.0, Opus 5.5) was used as follows. To check the manuscript and to improving writing. To prove the trapezoidal rule yields correct stationary states. 
To find the factor $(1-\epsilon/2)^{-N}$ that avoids potential overestimation when using the $\epsilon$ regularization.

\bibliography{references}

\onecolumngrid
\vspace{2\baselineskip}
\begin{center}
    \textbf{End Matter}
\end{center}
\vspace{\baselineskip}
\twocolumngrid

\emph{Comparison with Fermionic antiflatness}.---We compare our distance-based measure $\mathcal{D}$ and its lower bound $\mathcal{D}^{\mathrm{L}}$ with the recently introduced measure of fermionic antiflatness \cite{3yx4-1j27}, defined as, $ \mathcal{F}_1(|\Psi\rangle)=n-\frac{1}{2} \operatorname{tr}\left[\left(M^T M\right)\right]. $
Here, $M_{m n}=-i / 2\langle\psi|\left[\gamma_m, \gamma_n\right]|\psi\rangle$ defines the covariance matrix and $n = 2N$ are the number of fermionic modes. This measure bounds the distance-based measure \cite{haug2026practicaltestswitnessesfermionic} as \begin{equation}
    \frac{\mathcal{F}_1(|\psi\rangle)}{4 n} \leq \mathcal{D}(|\psi\rangle) \leq \frac{\mathcal{F}_1(|\psi\rangle)}{2}. 
\end{equation}

\begin{figure}
    \centering
    \includegraphics[width=0.9\linewidth]{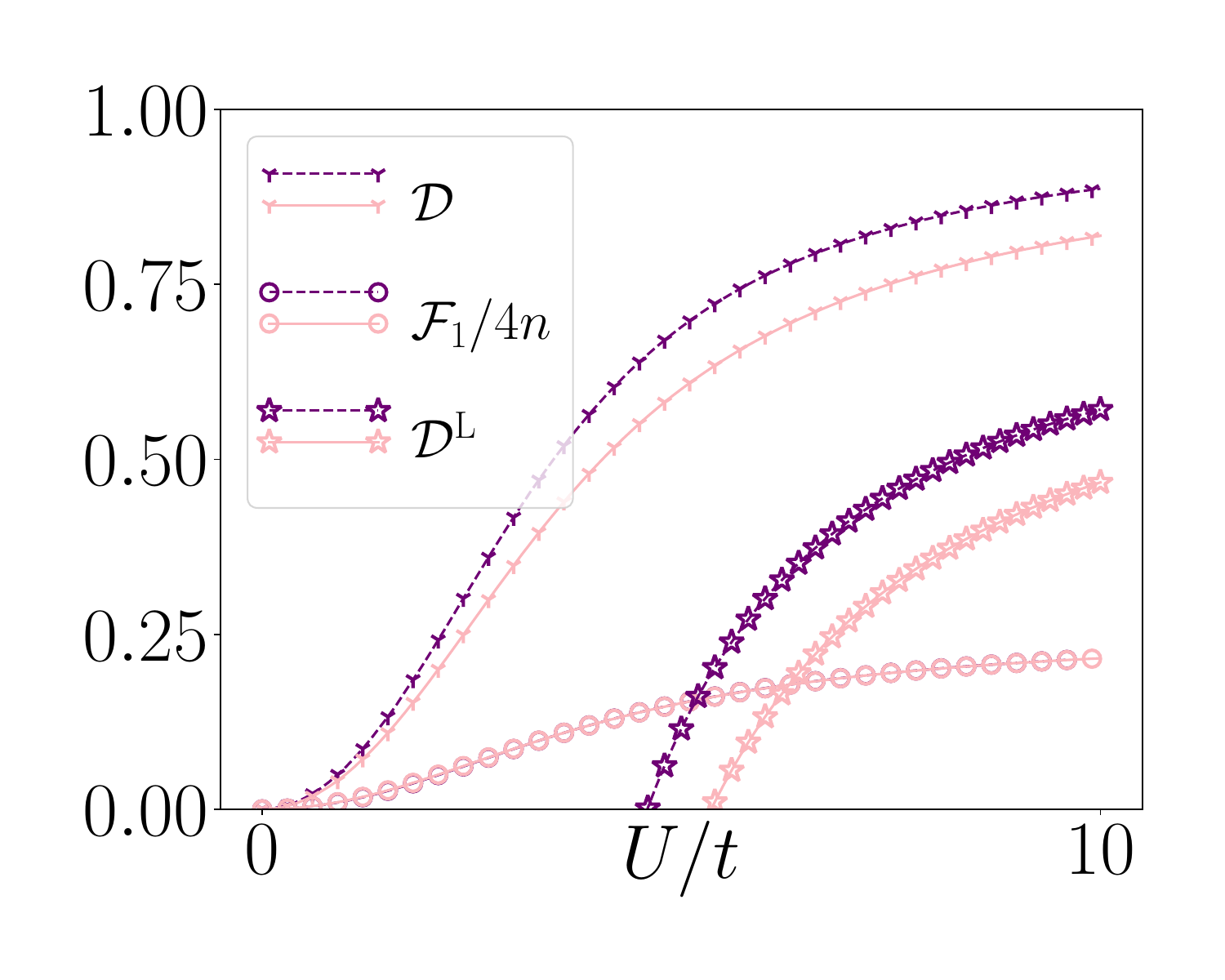}
    \caption{Lower bounds on the non-Gaussianity of the Fermi--Hubbard ground state at half
filling, for $N=8$ (light) and $N=10$ (dark) sites. Shown are the geometric measure
$\mathcal{D}(|\psi\rangle)$ obtained from the iterative procedure of
Theorem~\ref{thm:stationarity} (crosses), the lower bound
$\mathcal{D}^{\mathrm{L}}$ of Theorem~\ref{thm:lower-bound} (stars), and the antiflatness bound
$\mathcal{F}_1(|\psi\rangle)/4n$ (circles), with $n=2N$ the number of fermionic modes.
Below $U/t\approx5.0$ for $N=10$ and $U/t\approx6.2$ for $N=8$,
$\mathcal{D}^{\mathrm{L}}$ vanishes and antiflatness supplies the only
non-trivial bound; beyond these thresholds the ordering reverses. Antiflatness cannot
exceed $1/4$ by construction and is almost independent of system size; the two circle
curves nearly coincide, whereas $\mathcal{D}^{\mathrm{L}}$ keeps tracking the
growth of $\mathcal{D}$ and sharpens with $N$, its threshold moving to weaker
interactions as the system grows.}
    \label{fig:faf_vs_d}
\end{figure}

In Fig.~\ref{fig:faf_vs_d}, we compare the above lower bound with our lower bound $\mathcal{D}^\mathrm{L}$. We find the two to be complementary. At weak coupling, the non-Gaussian content is small, and antiflatness resolves it where our bound does not: $\mathcal{D}^{\mathrm{L}}$ becomes a genuine witness only once the variance term dominates. The ordering reverses, however, as $U/t$ grows: the $2$-norm of the covariance matrix associated with the half-filled ground state is driven towards zero, and  
$\mathcal{F}_1$ approaches its algebraic maximum and flattens. Intuitively, one might expect this behavior, since $\mathcal{F}_1$ does not account for higher-order correlations. In contrast, $\mathcal{D}^\mathrm{L}$ continues to track the growth of the exact measure up to the strongest couplings considered. These two witnesses to the distance-based monotone are therefore complementary; antiflatness works well at low interactions while the witness constructed in the present paper is tighter in the strongly-interacting regime. 

\emph{Proof of Theorem~\ref{thm:jensen-main}.}---Jensen's
inequality~\cite{jensen_original} states that $f(\mathbb{E}[x])\le
\mathbb{E}[f(x)]$ for a convex function $f$. Its quantum counterpart for a
bounded self-adjoint operator $A$ and a density operator $\rho$ reads
$f(\mu_A)\le\mathrm{tr}[f(A)\rho]$, with
$\mu_A=\mathrm{tr}(A\rho)$~\cite{HANSEN_PEDERSEN_2003}. This tells us only
that the Jensen gap, $\mathrm{tr}[f(A)\rho]-f(\mu_A)$, is non-negative; it
says nothing about how large the gap is. We aim to bound the gap from
both sides as
\begin{align*}
    h(\lambda_{\min};\mu_A)\operatorname{Var}_\rho(A) &\le 
    \mathrm{tr}[f(A)\rho]-f(\mu_A) \\ &\le
    h(\lambda_{\max};\mu_A)\operatorname{Var}_\rho(A),
\end{align*}
which for $\rho=\ket{\psi}\bra{\psi}$ is Eq.~\eqref{eq:jensen}. We proceed in
three steps.

\emph{Step 1: mean and variance as integrals over the spectrum.} By the
spectral theorem, $A=\int_{\sigma(A)}\lambda\,\mathrm{d}E(\lambda)$, where
$E$ is the projection-valued measure of $A$ on the Borel subsets of its
spectrum $\sigma(A)\subseteq[\lambda_{\min},\lambda_{\max}]$, with
$E(\sigma(A))=\mathbb{I}$ and $E(\Delta)E(\Delta')=E(\Delta\cap\Delta')$.
Functions of $A$ are defined through the functional calculus,
$g(A)=\int_{\sigma(A)}g(\lambda)\,\mathrm{d}E(\lambda)$, for any Borel
function $g$ bounded on $\sigma(A)$. The state $\rho$ turns $E$ into the
probability measure $\nu_\rho(\Delta)=\mathrm{tr}[\rho E(\Delta)]$ on
$\sigma(A)$---the distribution of outcomes when $A$ is measured in
$\rho$---so that
$\mathrm{tr}[g(A)\rho]=\int_{\sigma(A)}g(\lambda)\,\mathrm{d}\nu_\rho(\lambda)$.
Mean and variance thus take the form of ordinary averages,
\begin{equation}
    \mu_A=\int_{\sigma(A)}\lambda\,\mathrm{d}\nu_\rho(\lambda),\quad
    \operatorname{Var}_\rho(A)=\int_{\sigma(A)}(\lambda-\mu_A)^{2}\,
    \mathrm{d}\nu_\rho(\lambda).
    \label{eq:em-moments}
\end{equation}

\emph{Step 2: an exact identity for the scalar function.} For $f$ twice
differentiable on an interval containing $\sigma(A)$, the definition
$h(x;\mu)=[f(x)-f(\mu)]/(x-\mu)^{2}-f'(\mu)/(x-\mu)$ can be rearranged into
\begin{equation}
    f(x)=f(\mu)+f'(\mu)(x-\mu)+h(x;\mu)\,(x-\mu)^{2},
    \label{eq:em-identity}
\end{equation}
which holds for every $x$ in that interval, with $h(\mu;\mu)=f''(\mu)/2$ by
continuity. We stress that Eq.~\eqref{eq:em-identity} is not a truncated
Taylor expansion: all higher orders are contained in $h$, so the identity is
exact. This property is at the heart of the witness being non-perturbative. 
Moreover, $h(\cdot\,;\mu)$ is continuous and hence bounded on the
compact set $\sigma(A)$, so the functional calculus applies to it.

\emph{Step 3: lifting the identity to the operator.} Since
$f(A)=\int_{\sigma(A)}f(\lambda)\,\mathrm{d}E(\lambda)$, we may insert
Eq.~\eqref{eq:em-identity} under the integral. Using
$\int_{\sigma(A)}\mathrm{d}E(\lambda)=\mathbb{I}$ and
$\int_{\sigma(A)}\lambda\,\mathrm{d}E(\lambda)=A$, this gives
\begin{equation}
    f(A)=f(\mu)\,\mathbb{I}+f'(\mu)\,(A-\mu\mathbb{I})
    +\int_{\sigma(A)}h(\lambda;\mu)\,(\lambda-\mu)^{2}\,\mathrm{d}E(\lambda).
\end{equation}
Choosing $\mu=\mu_A$ and taking the expectation value in $\rho$, the linear
term vanishes, because $\mathrm{tr}[(A-\mu_A\mathbb{I})\rho]=0$. What
remains is an exact expression for the Jensen gap,
\begin{equation}
    \mathrm{tr}[f(A)\rho]-f(\mu_A)
    =\int_{\sigma(A)}h(\lambda;\mu_A)\,(\lambda-\mu_A)^{2}\,
    \mathrm{d}\nu_\rho(\lambda).
    \label{eq:em-gap}
\end{equation}

\emph{Bounding the gap.} Equation~\eqref{eq:em-gap} expresses the gap as an
average of $h(\lambda;\mu_A)$ with the weight
$(\lambda-\mu_A)^{2}\,\mathrm{d}\nu_\rho(\lambda)$. This weight is
non-negative and, by Eq.~\eqref{eq:em-moments}, integrates to
$\operatorname{Var}_\rho(A)$. Since $h(\cdot\,;\mu_A)$ is continuous on the
compact set $\sigma(A)$, it attains its minimum and maximum there; replacing
$h$ by either one under the integral leaves exactly
$\operatorname{Var}_\rho(A)$, and
\begin{align}
    \min_{\lambda\in\sigma(A)}h(\lambda;\mu_A)\,\operatorname{Var}_\rho(A)
    &\le\mathrm{tr}[f(A)\rho]-f(\mu_A) \nonumber \\ &\le
    \max_{\lambda\in\sigma(A)}h(\lambda;\mu_A)\,\operatorname{Var}_\rho(A).
    \label{eq:em-minmax}
\end{align}
If $h(\cdot\,;\mu_A)$ is non-decreasing on
$[\lambda_{\min},\lambda_{\max}]$, the minimum and maximum are attained at
$\lambda_{\min}$ and $\lambda_{\max}$, both of which belong to $\sigma(A)$
since the spectrum of a bounded self-adjoint operator is closed. Then
Eq.~\eqref{eq:em-minmax} becomes Eq.~\eqref{eq:jensen}, which completes the
proof.

\emph{Monotonicity for $f=\exp$.} The function $h$ is non-decreasing
whenever $f'$ is convex~\cite{liao2019sharpening}, which covers $f=\exp$,
the case used in this work. For completeness, we verify it directly. Writing
$u=x-\mu$, one has $h(x;\mu)=e^{\mu}\varphi(u)$ with
$\varphi(u)=(e^{u}-1-u)/u^{2}$, whose derivative is
$\varphi'(u)=\nu(u)/u^{3}$ with $\nu(u)=(u-2)e^{u}+u+2$. The function $\nu$
satisfies $\nu(0)=\nu'(0)=0$ and $\nu''(u)=ue^{u}$, which has the sign of
$u$. Hence $\nu$ has the same sign as $u^{3}$, so that $\varphi'(u)>0$ for
all $u\neq0$, and $h$ is strictly increasing.

The scalar version of Eq.~\eqref{eq:em-gap} appears in
Ref.~\cite{liao2019sharpening}; the spectral theorem carries it over to
operators, and the monotonicity of $h$ reduces the bounds to an evaluation
at the two spectral edges.

% ===================== SUPPLEMENTAL MATERIAL =====================
\clearpage
\onecolumngrid
\let\addcontentsline\ORIGaddcontentsline  % restore
\setcounter{secnumdepth}{3}               % REVTeX letters run at 0 -> no S1, S2

\setcounter{equation}{0}
\setcounter{figure}{0}
\setcounter{table}{0}
\setcounter{section}{0}
\setcounter{theorem}{0}
\setcounter{proposition}{0}

\renewcommand{\theequation}{S\arabic{equation}}
\renewcommand{\thefigure}{S\arabic{figure}}
\renewcommand{\thetable}{S\arabic{table}}
\renewcommand{\thesection}{S\arabic{section}}
\renewcommand{\thesubsection}{S\arabic{section}.\arabic{subsection}}
\renewcommand{\thetheorem}{S\arabic{theorem}}
\renewcommand{\theproposition}{S\arabic{proposition}}

\begin{center}
  \textbf{\large Supplemental Material}\\[4pt]
  \textbf{Fermionic non-Gaussianity via computable measures and
          experimentally accessible bounds}\\[4pt]
  Leela Ganesh Chandra Lakkaraju, Naga Dileep Varikuti, and Philipp Hauke
\end{center}
\vspace{6pt}

\addtocontents{toc}{\protect\setcounter{tocdepth}{1}}   
\setcounter{tocdepth}{2}
\vspace{-1.5em}
\tableofcontents
\vspace{1em}

\section{Monotonicity of the fidelity-based measure}
\label{sec:monotonicity}

Throughout, $\mathcal{G}$ denotes the set of fermionic Gaussian states, taken closed so
that it contains the pure Gaussian states, and
$\mathcal{F}(\rho,\sigma)=\big(\mathrm{Tr}\sqrt{\sqrt{\rho}\,\sigma\sqrt{\rho}}\big)^{2}$
is the Uhlmann--Jozsa fidelity in the convention of the main text. We abbreviate the
Gaussian fidelity and the measure as
\begin{equation}
    \mathcal{F}_{\mathcal{G}}(\rho) \;=\; \max_{\sigma\in\mathcal{G}}
    \mathcal{F}(\rho,\sigma), \qquad
    \mathcal{D}(\rho) \;=\; 1-\mathcal{F}_{\mathcal{G}}(\rho),
    \label{eq:Dmixed}
\end{equation}
which is Eq.~(1) of the main text. The maximum is attained: Gaussian states are
parametrised by covariance matrices with $|m_k|\le 1$, a compact set, and
$\mathcal{F}(\rho,\cdot)$ is continuous on it. Both properties below hold for arbitrary,
possibly mixed $\rho$; the restriction to pure states in the main text is one of physical
interest, not of validity. Faithfulness is proved in the main text, and we establish here
the two remaining properties of Proposition~1.

\subsection{Invariance under Gaussian unitaries}

A Gaussian unitary is an automorphism of the Gaussian set,
$U_{G}\,\mathcal{G}\,U_{G}^{\dagger}=\mathcal{G}$. Unitary invariance of the fidelity gives
\begin{align}
    \mathcal{F}_{\mathcal{G}}\big(U_{G}\rho U_{G}^{\dagger}\big)
    &= \max_{\sigma\in\mathcal{G}}
       \mathcal{F}\big(U_{G}\rho U_{G}^{\dagger},\,\sigma\big) \nonumber \\
    &= \max_{\sigma\in\mathcal{G}}
       \mathcal{F}\big(\rho,\,U_{G}^{\dagger}\sigma U_{G}\big)
     = \mathcal{F}_{\mathcal{G}}(\rho),
    \label{eq:unitary}
\end{align}
the last equality because $\sigma'=U_{G}^{\dagger}\sigma U_{G}$ ranges over all of
$\mathcal{G}$ as $\sigma$ does. Hence
$\mathcal{D}(U_{G}\rho U_{G}^{\dagger})=\mathcal{D}(\rho)$.

\subsection{Monotonicity under Gaussian channels}

Let $\mathcal{E}_{G}$ be a Gaussian CPTP map, that is,
$\mathcal{E}_{G}(\mathcal{G})\subseteq\mathcal{G}$. The fidelity obeys the data-processing
inequality~\cite{NielsenChuang},
\begin{equation}
    \mathcal{F}\big(\mathcal{E}(\rho),\mathcal{E}(\sigma)\big)
    \;\ge\; \mathcal{F}(\rho,\sigma)
    \qquad \text{for every CPTP } \mathcal{E}.
    \label{eq:dpi}
\end{equation}
Fix any $\sigma\in\mathcal{G}$. Since $\mathcal{E}_{G}(\sigma)\in\mathcal{G}$, it is an
admissible competitor in the maximisation defining
$\mathcal{F}_{\mathcal{G}}(\mathcal{E}_{G}(\rho))$, so that
\begin{equation}
    \mathcal{F}_{\mathcal{G}}\big(\mathcal{E}_{G}(\rho)\big)
    \;\ge\; \mathcal{F}\big(\mathcal{E}_{G}(\rho),\mathcal{E}_{G}(\sigma)\big)
    \;\ge\; \mathcal{F}(\rho,\sigma).
    \label{eq:chain}
\end{equation}
The left-hand side carries no $\sigma$ dependence. Maximising the right-hand side over
$\sigma\in\mathcal{G}$ therefore gives
$\mathcal{F}_{\mathcal{G}}(\mathcal{E}_{G}(\rho))\ge\mathcal{F}_{\mathcal{G}}(\rho)$, and
with Eq.~\eqref{eq:Dmixed},
\begin{equation}
    \mathcal{D}\big(\mathcal{E}_{G}(\rho)\big) \;\le\; \mathcal{D}(\rho).
    \label{eq:monotone}
\end{equation}
The two inequalities in Eq.~\eqref{eq:chain} have different origins: the first holds
because $\mathcal{E}_{G}$ preserves the free set, the second because the channel acts on
both arguments at once. Neither step requires $\rho$ to be pure, and neither requires
$\mathcal{G}$ to be convex.

\section{The ordered average and its trapezoidal evaluation}
\label{sm:ordered}

\subsection{Conventions and the canonical form}

We use the $2N$ Majorana operators $\gamma_{2j-1}=c_j+c^\dagger_j$ and
$\gamma_{2j}=-i(c_j-c^\dagger_j)$, which are Hermitian, square to unity, and
obey $\{\gamma_a,\gamma_b\}=2\delta_{ab}$. For $a\neq b$ we abbreviate
$X_{ab}=i\gamma_a\gamma_b$, Hermitian with $X_{ab}^2=\mathbb{I}$ and
$X_{ba}=-X_{ab}$. The Gaussian state is
$\sigma_K=e^{-\Gamma}/\mathcal{Z}_K$ with
\begin{equation}
    \Gamma=\frac{i}{4}\sum_{a,b}K_{ab}\gamma_a\gamma_b
          =\frac{1}{2}\sum_{a<b}K_{ab}X_{ab},
    \qquad
    \frac{\partial\Gamma}{\partial K_{ab}}=\frac{1}{2}X_{ab}
    \quad(a<b),
    \label{sm:eq:gamma}
\end{equation}
and covariance matrix $M_{ab}=\operatorname{Tr}(\sigma_KX_{ab})$, so that
$\langle X_{ab}\rangle_{\sigma_K}=M_{ab}$ exactly. Only the antisymmetric
part of the coefficient matrix survives in Eq.~\eqref{sm:eq:gamma}, since
$\gamma_a\gamma_b=-\gamma_b\gamma_a$ for $a\neq b$, and the factor $i$ makes
$\Gamma$ Hermitian; we take $K$ real and antisymmetric throughout.

A real antisymmetric matrix is brought by a real orthogonal transformation
to $2\times2$ blocks,
\begin{equation}
    K=O\Big(\bigoplus_k\beta_k\varepsilon\Big)O^{\mathsf T},
    \qquad
    \varepsilon=\begin{pmatrix}0&1\\-1&0\end{pmatrix},
    \label{sm:eq:youla}
\end{equation}
and the same $O$ is absorbed into the operators. Setting
$\tilde\gamma_c=\sum_aO_{ac}\gamma_a$, orthogonality of $O$ gives
$\{\tilde\gamma_c,\tilde\gamma_d\}=2\delta_{cd}$, so the rotated operators
are again Majoranas. Each block then contributes
$\tfrac{i}{2}\beta_k\tilde\gamma_{2k-1}\tilde\gamma_{2k}$, and with the
Bogoliubov fermion $d_k=(\tilde\gamma_{2k-1}+i\tilde\gamma_{2k})/2$ and
$i\tilde\gamma_{2k-1}\tilde\gamma_{2k}=2n_k-1$,
\begin{equation}
    \Gamma=\sum_k\beta_k\Big(n_k-\frac{1}{2}\Big),
    \qquad
    \mathcal{Z}_K=\prod_k2\cosh(\beta_k/2),
    \qquad
    \langle n_k\rangle=\frac{1}{e^{\beta_k}+1}.
    \label{sm:eq:canonical}
\end{equation}
The covariance matrix is brought to the same block form by $O$, with
\begin{equation}
    m_k=\langle 2n_k-1\rangle=-\tanh(\beta_k/2),
    \qquad
    |\beta_k|=2\operatorname{artanh}|m_k|
             =\ln\frac{1+|m_k|}{1-|m_k|}.
    \label{sm:eq:mk}
\end{equation}
We write $\beta_{\max}=\max_k|\beta_k|=2\operatorname{artanh}|m|_{\max}$.
The same frame gives $\lambda_{\max}(-\Gamma)=\tfrac12\sum_k|\beta_k|$,
attained on the Bogoliubov configuration with $n_k=0$ for $\beta_k>0$ and
$n_k=1$ for $\beta_k<0$, itself a pure Gaussian state. Note that
$\beta_{\max}$ is set by how close the current Gaussian state sits to the
pure-state boundary $|m_k|=1$, and diverges logarithmically as that boundary
is approached.

\subsection{Derivative of the fidelity}

Given a differentiable family of bounded operators, the derivative of the
exponential is an ordered average along it, given by
\begin{equation}
    \frac{\partial}{\partial g}e^{-\Gamma}
    =-\int_0^1\!\mathrm{d}s\;
      e^{-s\Gamma}\Big(\frac{\partial\Gamma}{\partial g}\Big)e^{-(1-s)\Gamma},
    \label{sm:eq:duhamel}
\end{equation}
which reduces to $-(\partial_g\Gamma)e^{-\Gamma}$ only when
$[\Gamma,\partial_g\Gamma]=0$. Everything below follows from the fact that
Eq.~\eqref{sm:eq:duhamel} collapses under a trace but not inside a
pure-state expectation value.

Under the trace, cyclicity recombines $e^{-s\Gamma}$ with
$e^{-(1-s)\Gamma}$ for every $s$ and the remaining integral is that of a
constant,
\begin{equation}
    \frac{\partial\mathcal{Z}_K}{\partial K_{ab}}
    =-\frac{1}{2}\operatorname{Tr}\big(X_{ab}e^{-\Gamma}\big)
    =-\frac{1}{2}\mathcal{Z}_K M_{ab}.
    \label{sm:eq:dZ}
\end{equation}
No such recombination is available in
$A=\bra{\psi}e^{-\Gamma}\ket{\psi}$. Writing
\begin{equation}
    g_{ab}(s)=\bra{\psi}\,e^{-s\Gamma}\,X_{ab}\,e^{-(1-s)\Gamma}\,\ket{\psi},
    \qquad
    N_{ab}=\int_0^1\!\mathrm{d}s\;g_{ab}(s),
    \label{sm:eq:g}
\end{equation}
one has $\partial A/\partial K_{ab}=-\tfrac12N_{ab}$, and therefore
\begin{equation}
    \frac{\partial\ln\mathcal{F}}{\partial K_{ab}}
    =\frac{1}{2}\Big(M_{ab}-\frac{N_{ab}}{A}\Big),
    \label{sm:eq:gradient}
\end{equation}
whose zeros are Eq.~(\ref{eq:stationarity}) of the main text. Only $N_{ab}$
is quadratured. The objective $\mathcal{F}$, the covariance matrix $M$ and
the normalisation $A$ are computed exactly, so the value of the fidelity at
any $K$ carries no quadrature error whatever, regardless of how that $K$ was
reached. We have checked Eq.~\eqref{sm:eq:gradient} against central finite
differences of $\mathcal{F}(K)$ in a Jordan--Wigner representation with
$N=3$ modes, finding agreement to $10^{-10}$.

\subsection{Real-valuedness of the ordered average}

Since $\Gamma$ and $X_{ab}$ are Hermitian,
$g_{ab}(s)^{*}=g_{ab}(1-s)$. The ordered average is therefore real, as it
must be because $M$ is real, and the trapezoidal rule collapses to a single
endpoint,
\begin{equation}
    N^{\mathrm{trap}}_{ab}=\frac{g_{ab}(0)+g_{ab}(1)}{2}
    =\operatorname{Re}\bra{\psi}X_{ab}e^{-\Gamma}\ket{\psi},
    \label{sm:eq:trap}
\end{equation}
which is Eq.~(\ref{eq:update}) of the main text. Antisymmetry,
$N_{ba}=-N_{ab}$, is preserved because $X_{ba}=-X_{ab}$. In the
particle-conserving sector this is purely algebraic: there
$\big(\bra{\psi}c^\dagger_jc_ke^{-\Gamma}\ket{\psi}\big)^{*}
 =\bra{\psi}e^{-\Gamma}c^\dagger_kc_j\ket{\psi}$, so taking the Hermitian
part of the update is exactly the average of its two endpoints. An
implementation that symmetrises $C$ is therefore already evaluating the
trapezoidal rule.

\subsection{The exact error kernel}

Let $\Gamma\ket{x}=E_x\ket{x}$, $\psi_x=\braket{x|\psi}$, and
$a_x=e^{-E_x}$, $\Delta=E_x-E_y$. Inserting resolutions of the identity in
Eq.~\eqref{sm:eq:g}, both numerators are the same bilinear form with
different kernels,
\begin{equation}
    w^{\mathrm{ex}}_{xy}=\int_0^1\!\mathrm{d}s\,a_x^{s}a_y^{1-s}
    =\frac{a_y-a_x}{\ln a_y-\ln a_x},
    \qquad
    w^{\mathrm{trap}}_{xy}=\frac{a_x+a_y}{2},
    \label{sm:eq:weights}
\end{equation}
the logarithmic and the arithmetic mean of the two Boltzmann factors. Their
ratio is universal,
\begin{equation}
    \frac{w^{\mathrm{trap}}_{xy}}{w^{\mathrm{ex}}_{xy}}=r(\Delta)
    =\frac{\Delta}{2}\coth\frac{\Delta}{2}
    =1+\frac{\Delta^2}{12}-\frac{\Delta^4}{720}+\mathcal{O}(\Delta^6)\ \ge\ 1,
    \label{sm:eq:ratio}
\end{equation}
even, real-analytic, equal to unity at $\Delta=0$ and strictly increasing in
$|\Delta|$. The distortion is thus one-signed---the trapezoidal rule always
overweights---and depends on the spectral pair only through $\Delta$. That
last property is what makes the argument of the next subsection possible.

Which $\Delta$ occur is fixed by the algebra. Each rotated Majorana is
linear in the Bogoliubov operators, so $X_{ab}$ changes at most two
occupations, through the channels $\mathbb{I},n_k$ ($\Delta=0$),
$d^\dagger_kd_l$ ($\Delta=\beta_k-\beta_l$), and $d^\dagger_kd^\dagger_l$,
$d_kd_l$ ($\Delta=\pm(\beta_k+\beta_l)$). Hence
\begin{equation}
    |\Delta|\le2\beta_{\max},
    \qquad
    1\le r\le\beta_{\max}\coth\beta_{\max},
    \label{sm:eq:kernelbound}
\end{equation}
independently of $N$: a bilinear probes the two-quasiparticle scale
$2\beta_{\max}$ and not the extensive many-body width $\sum_k|\beta_k|$.

Equation~\eqref{sm:eq:kernelbound} bounds each kernel, and it does not
transfer to the summed update as a relative bound, because the coefficients
$\bar\psi_x\psi_y(X_{ab})_{xy}$ carry signs and phases and an entry of
$N_{ab}$ can pass through zero. What survives summation is an absolute
$\ell^1$ bound and a relative bound inflated by a computable cancellation
number. We stress, however, that this gap is not the operative issue, and
neither is the size of Eq.~\eqref{sm:eq:kernelbound}. At the regularised
optimum the iteration runs at $\beta_{\max}\simeq\ln(2/\epsilon)$, so that
$\epsilon=10^{-6}$ already gives $\beta_{\max}\simeq14.5$ and a permitted
kernel distortion above $10^3\%$. Any argument for correctness resting on
smallness would fail there. The correct statement is stronger and does not
involve smallness at all.

\subsection{Root-space diagonality: the fixed points coincide}

Let $\mathfrak{g}=\operatorname{span}_{\mathbb C}\{X_{ab}\}_{a<b}\cong
\mathfrak{so}(2N)_{\mathbb C}$. Then $\mathfrak{g}$ is closed under
$\mathrm{ad}_\Gamma(\cdot)=[\Gamma,\cdot]$, and $\mathrm{ad}_\Gamma$ is
diagonalisable on it with real spectrum
\begin{equation}
    \operatorname{spec}\mathrm{ad}_\Gamma
    =\{0\}\cup\{\pm(\beta_k-\beta_l),\ \pm(\beta_k+\beta_l)\}_{k<l},
    \label{sm:eq:spectrum}
\end{equation}
read off from the four channels above. The zero eigenspace is the Cartan
subalgebra $\mathfrak{h}=\operatorname{span}\{2n_k-1\}$, of dimension $N$.
Define, for $Y\in\mathfrak{g}$, the exact and trapezoidal gradient
functionals
\begin{equation}
    \mathcal{D}[Y]=\langle Y\rangle_{\sigma_K}
      -\frac{1}{A}\int_0^1\!\mathrm{d}s\,
       \bra{\psi}e^{-s\Gamma}Ye^{-(1-s)\Gamma}\ket{\psi},
    \qquad
    \mathcal{D}_{\mathrm t}[Y]=\langle Y\rangle_{\sigma_K}
      -\frac{1}{2A}\Big(\bra{\psi}Ye^{-\Gamma}\ket{\psi}
                       +\bra{\psi}e^{-\Gamma}Y\ket{\psi}\Big),
    \label{sm:eq:functionals}
\end{equation}
so that $\mathcal{D}[X_{ab}]=2\,\partial\ln\mathcal{F}/\partial K_{ab}$ and
$\mathcal{D}_{\mathrm t}$ is its trapezoidal surrogate.

\begin{theorem}
\label{sm:thm:rootspace}
Set $\varphi(\lambda)=\int_0^1e^{-s\lambda}\mathrm{d}s=(1-e^{-\lambda})/\lambda$
and $\tau(\lambda)=(1+e^{-\lambda})/2$, both strictly positive for every
real $\lambda$, with $\varphi(0)=\tau(0)=1$ and
$\tau(\lambda)/\varphi(\lambda)=r(\lambda)$. Then
$\mathcal{D}_{\mathrm t}=\mathcal{P}\mathcal{D}$, where
$\mathcal{P}=\mathrm{id}_{\mathfrak h}\oplus\bigoplus_\alpha r(\alpha(\beta))$
is positive-definite and diagonal in the root decomposition, with
\begin{equation}
    1\le\operatorname{spec}\mathcal{P}\le\beta_{\max}\coth\beta_{\max}.
\end{equation}
In particular $\mathcal{D}=0\iff\mathcal{D}_{\mathrm t}=0$: the exact map and
the trapezoidal map have identical fixed points.
\end{theorem}

\begin{proof}
Let $Y\in\mathfrak{g}_\alpha$, so $\mathrm{ad}_\Gamma Y=\lambda Y$ and
$e^{-s\Gamma}Ye^{s\Gamma}=e^{-s\,\mathrm{ad}_\Gamma}Y=e^{-s\lambda}Y$. Then
\begin{equation}
    e^{-s\Gamma}Ye^{-(1-s)\Gamma}
    =\big(e^{-s\Gamma}Ye^{s\Gamma}\big)e^{-\Gamma}
    =e^{-s\lambda}\,Ye^{-\Gamma},
    \label{sm:eq:factorisation}
\end{equation}
an exact factorisation of the $s$-dependence, valid whether or not
$[\Gamma,Y]=0$. With $h_Y=\bra{\psi}Ye^{-\Gamma}\ket{\psi}$, integrating
gives $\varphi(\lambda)h_Y$, while averaging the endpoints gives
$\tau(\lambda)h_Y$: at $s=0$ the integrand is $h_Y$, and at $s=1$ it is
$e^{-\lambda}h_Y$. For $\lambda\neq0$, cyclicity together with
Eq.~\eqref{sm:eq:factorisation} gives
$\operatorname{Tr}(Ye^{-\Gamma})=e^{-s\lambda}\operatorname{Tr}(Ye^{-\Gamma})$
for all $s$, hence $\langle Y\rangle_{\sigma_K}=0$, and
$\mathcal{D}[Y]=-\varphi(\lambda)h_Y/A$,
$\mathcal{D}_{\mathrm t}[Y]=-\tau(\lambda)h_Y/A$, so
$\mathcal{D}_{\mathrm t}[Y]=r(\lambda)\mathcal{D}[Y]$. For $Y\in\mathfrak{h}$,
$\lambda=0$ and $\varphi=\tau=1$, so the two coincide identically. Both
functionals are linear on $\mathfrak{g}$, on which the root spaces form a
basis; since $\varphi,\tau>0$ everywhere, they have the same kernel. The
bound on $\operatorname{spec}\mathcal{P}$ is Eq.~\eqref{sm:eq:kernelbound}.
\end{proof}

The trapezoidal rule is therefore not an approximation of the answer but a
preconditioner of the iteration. Three consequences follow. The Cartan
sector $\{m_k\}$, which is the polarisation of each canonical mode, receives
no quadrature error at all; this is the general form of the statement that
diagonal elements are exact because $[\Gamma,n_k]=0$, and it holds without
any assumption that $\Gamma$ and $X_{ab}$ commute. The root sector, which
rotates the canonical frame, is systematically over-rotated by $r(\lambda)$,
at most $\beta_{\max}\coth\beta_{\max}$; damping or under-relaxation by
$1/r$ is therefore the correct compensation rather than a cosmetic device.
And the fixed point, the value of $\mathcal{F}$ there, and hence the reported
non-Gaussianity are unaffected.

Equation~\eqref{sm:eq:factorisation} also shows that $g_{ab}(s)$ is a sum of
at most $N(2N-1)+1$ pure exponentials with known rates $|\lambda|\le
2\beta_{\max}$. The quadrature error is consequently not merely bounded but
computable, $N^{\mathrm{trap}}_{ab}-N_{ab}=\sum_\lambda[\tau(\lambda)
-\varphi(\lambda)]h^{(ab)}_\lambda$, and removable by dividing each root
component by $r(\lambda)$. Gauss--Legendre quadrature converges
superexponentially on such an integrand, and
$n\simeq\lceil\beta_{\max}\rceil+4$ nodes reach machine precision. Should
the exact trajectory be wanted rather than only the exact fixed point, this
is the remedy, at the cost of $n$ evaluations instead of two.

Finally, the scheme is correct in total for reasons that do not depend on
any of this. Every $K$ gives
$\mathcal{F}(K)=\bra{\psi}\sigma_K\ket{\psi}\le\max_{\sigma \in \mathcal{G}}
\bra{\psi}\sigma\ket{\psi}$, evaluated without quadrature, so any iterate---%
exact, trapezoidal, or stopped early---returns a rigorous upper bound on
$\mathcal{D}$. Quadrature can cost tightness; it cannot cost validity. What
Theorem~\ref{sm:thm:rootspace} adds is sharpness: the iteration converges to
a genuine stationary point of the exact objective, not to a nearby
surrogate.

\section{Stationarity of the bound}
\label{sm:bound}

\subsection{Setup}

We keep the conventions of Sec.~\ref{sm:ordered}: $\Gamma$ is given by
Eq.~\eqref{sm:eq:gamma}, $\partial\Gamma/\partial K_{ab}=X_{ab}/2$ for
$a<b$, and $M_{ab}=\operatorname{Tr}(\sigma_KX_{ab})$ is the covariance
matrix of the trial state. Throughout we work with
\begin{equation}
    \mu=\bra{\psi}(-\Gamma)\ket{\psi},
    \quad
    V=\operatorname{Var}_\psi(\Gamma),
    \quad
    \lambda_{\max}=\lambda_{\max}(-\Gamma),
    \quad
    \delta=\lambda_{\max}-\mu,
    \label{sm:eq:bound-notation}
\end{equation}
and abbreviate $h=h(\lambda_{\max};\mu)$ and
$\alpha=e^{\mu}+hV$, so that $B(K)=\alpha/\mathcal{Z}_K$. Note that
$\operatorname{Var}_\psi(\Gamma)=\operatorname{Var}_\psi(-\Gamma)$, so no
sign ambiguity arises in $V$; the sign convention matters only in $\mu$,
which is the mean of $-\Gamma$ and \emph{not} of $\Gamma$. This is the one
place in the derivation where a sign error propagates into every term, and
we flag it explicitly.

For $f=\exp$ the second divided difference and its partial derivatives are
\begin{align}
    h(\lambda;\mu)&=\frac{e^{\lambda}-e^{\mu}}{\delta^{2}}-\frac{e^{\mu}}{\delta},
    \label{sm:eq:h}\\[2pt]
    \partial_\lambda h&=\frac{e^{\lambda}+e^{\mu}}{\delta^{2}}
                        -\frac{2(e^{\lambda}-e^{\mu})}{\delta^{3}},
    \label{sm:eq:dh-lam}\\[2pt]
    \partial_\mu h&=-\frac{e^{\mu}}{\delta}-\frac{2e^{\mu}}{\delta^{2}}
                    +\frac{2(e^{\lambda}-e^{\mu})}{\delta^{3}}.
    \label{sm:eq:dh-mu}
\end{align}
Both follow by differentiating Eq.~\eqref{sm:eq:h} at fixed $\mu$ and fixed
$\lambda$ respectively, using $\partial\delta/\partial\lambda=1$ and
$\partial\delta/\partial\mu=-1$.

\subsection{Monotonicity of \texorpdfstring{$h$}{h}, and why no spectral
optimisation is needed}

Writing $u=\lambda-\mu$, Eq.~\eqref{sm:eq:h} reads
$h=e^{\mu}\varphi(u)$ with $\varphi(u)=(e^{u}-1-u)/u^{2}$. Then
\begin{equation}
    \varphi'(u)=\frac{\nu(u)}{u^{3}},
    \qquad
    \nu(u)=u e^{u}+u-2e^{u}+2 .
\end{equation}
One has $\nu(0)=0$, $\nu'(u)=e^{u}(u-1)+1$ with $\nu'(0)=0$, and
$\nu''(u)=ue^{u}$. Hence $\nu''$ has the sign of $u$, so $\nu'$ attains its
minimum $0$ at $u=0$ and $\nu'\ge0$ everywhere; $\nu$ is therefore
non-decreasing with $\nu(0)=0$, so $\nu$ and $u^{3}$ share their sign and
$\varphi'>0$ for all $u\neq0$. Thus $h(\cdot\,;\mu)$ is strictly increasing.
Consequently
\begin{equation}
    h(\lambda_{\min};\mu)\le h(\lambda;\mu)\le h(\lambda_{\max};\mu)
    \quad\text{for all }\lambda\in\operatorname{spec}(-\Gamma),
\end{equation}
and Eq.~(\ref{eq:jensen}) of the main text follows from the exact expansion
$\bra{\psi}f(A)\ket{\psi}-f(\mu_A)=\sum_\lambda h(\lambda;\mu_A)
(\lambda-\mu_A)^{2}p_\lambda$ with $p_\lambda=|\braket{\lambda|\psi}|^{2}$,
by bounding $h$ above and below on the spectrum. No optimisation over the
spectrum is required, and the two edges give the tightest bounds of this
form.

The expressions~\eqref{sm:eq:h}--\eqref{sm:eq:dh-mu} have removable
singularities at $\delta=0$. From
$\varphi(u)=\tfrac12+\tfrac{u}{6}+\tfrac{u^{2}}{24}+\mathcal{O}(u^{3})$ and
$h=e^{\mu}\varphi$, $\partial_\lambda h=e^{\mu}\varphi'$,
$\partial_\mu h=e^{\mu}(\varphi-\varphi')$, one obtains the limits
\begin{equation}
    h\to\frac{e^{\mu}}{2},
    \qquad
    \partial_\lambda h\to\frac{e^{\mu}}{6},
    \qquad
    \partial_\mu h\to\frac{e^{\mu}}{3}
    \qquad(\delta\to0),
    \label{sm:eq:removable}
\end{equation}
which we evaluate from the series below a threshold on $|\delta|$.

\subsection{The core condition}

Differentiating $B=\alpha/\mathcal{Z}_K$ and using
Eq.~\eqref{sm:eq:dZ},
\begin{equation}
    \frac{\partial B}{\partial K_{ab}}
    =\frac{1}{\mathcal{Z}_K}\frac{\partial\alpha}{\partial K_{ab}}
     -\frac{\alpha}{\mathcal{Z}_K^{2}}\frac{\partial\mathcal{Z}_K}{\partial K_{ab}}
    =\frac{1}{\mathcal{Z}_K}\frac{\partial\alpha}{\partial K_{ab}}
     +\frac{\alpha}{2\mathcal{Z}_K}M_{ab}.
    \label{sm:eq:dB}
\end{equation}
Since $\alpha>0$ wherever the bound is meaningful, $B$ is stationary exactly
where
\begin{equation}
    M_{ab}=-\frac{2}{\alpha}\frac{\partial\alpha}{\partial K_{ab}} .
    \label{sm:eq:core}
\end{equation}
The factor $2$ is a consequence of the Majorana normalisation
$\partial\Gamma/\partial K_{ab}=X_{ab}/2$ and is absent in a Dirac-form
derivation. Everything that follows is the evaluation of
$\partial\alpha/\partial K_{ab}$ term by term.

\subsection{The four derivatives}

\paragraph{Mean.} With $\partial(-\Gamma)/\partial K_{ab}=-X_{ab}/2$ and
$\ket{\psi}$ fixed,
\begin{equation}
    \frac{\partial\mu}{\partial K_{ab}}
    =-\frac{1}{2}\bra{\psi}X_{ab}\ket{\psi}
    =-\frac{1}{2}M^{\psi}_{ab}.
    \label{sm:eq:dmu}
\end{equation}

\paragraph{Variance.} From $V=\bra{\psi}(-\Gamma)^{2}\ket{\psi}-\mu^{2}$,
the product rule inside the expectation value gives
$\partial_{K_{ab}}\bra{\psi}(-\Gamma)^{2}\ket{\psi}
=-\tfrac12\bra{\psi}\{-\Gamma,X_{ab}\}\ket{\psi}=-S_{ab}$ with $S$ as in
Eq.~(\ref{eq:S-tensor}) of the main text, while
$\partial_{K_{ab}}\mu^{2}=2\mu\,\partial_{K_{ab}}\mu=-\mu M^{\psi}_{ab}$.
Hence
\begin{equation}
    \frac{\partial V}{\partial K_{ab}}=-S_{ab}+\mu M^{\psi}_{ab}.
    \label{sm:eq:dV}
\end{equation}
Note that $S_{ab}=\operatorname{Re}\bra{\psi}(-\Gamma)X_{ab}\ket{\psi}$ is
real, and is quartic in the Majorana operators because $\Gamma$ is
quadratic and $X_{ab}$ bilinear.

\paragraph{Spectral edge.} By the Hellmann--Feynman theorem applied to
$-\Gamma$, with $\ket{\lambda_{\max}}$ its top eigenvector,
\begin{equation}
    \frac{\partial\lambda_{\max}}{\partial K_{ab}}
    =\bra{\lambda_{\max}}\frac{\partial(-\Gamma)}{\partial K_{ab}}\ket{\lambda_{\max}}
    =-\frac{1}{2}\bar{M}_{ab},
    \qquad
    \bar{M}_{ab}=\bra{\lambda_{\max}}X_{ab}\ket{\lambda_{\max}}.
    \label{sm:eq:dlam}
\end{equation}
This step requires $\lambda_{\max}$ to be non-degenerate; the degenerate
case is discussed at the end of this section.

\paragraph{Bounding function.} $h$ depends on $K$ through both arguments,
so by Eqs.~\eqref{sm:eq:dmu} and~\eqref{sm:eq:dlam},
\begin{equation}
    \frac{\partial h}{\partial K_{ab}}
    =-\frac{1}{2}\Big[(\partial_\lambda h)\,\bar{M}_{ab}
                     +(\partial_\mu h)\,M^{\psi}_{ab}\Big].
    \label{sm:eq:dh}
\end{equation}

\subsection{Assembly}

Collecting Eqs.~\eqref{sm:eq:dmu}, \eqref{sm:eq:dV} and~\eqref{sm:eq:dh} in
$\alpha=e^{\mu}+hV$,
\begin{equation}
\begin{aligned}
    \frac{\partial\alpha}{\partial K_{ab}}
    &=e^{\mu}\frac{\partial\mu}{\partial K_{ab}}
     +V\frac{\partial h}{\partial K_{ab}}
     +h\frac{\partial V}{\partial K_{ab}}\\[2pt]
    &=-\frac{e^{\mu}}{2}M^{\psi}_{ab}
      -\frac{V}{2}\Big[(\partial_\lambda h)\bar{M}_{ab}
                      +(\partial_\mu h)M^{\psi}_{ab}\Big]
      -hS_{ab}+h\mu M^{\psi}_{ab}.
\end{aligned}
\end{equation}
Substituting into Eq.~\eqref{sm:eq:core} and grouping the three tensors,
\begin{equation}
    -\frac{2}{\alpha}\frac{\partial\alpha}{\partial K_{ab}}
    =\big(c_0M^{\psi}+c_1\bar{M}+c_2S\big)_{ab},
\end{equation}
with $c_0,c_1,c_2$ exactly as in Theorem \ref{thm:bound-iteration} of the main
text. Reinserting this into Eq.~\eqref{sm:eq:dB} gives the gradient in the
form quoted as Eq.~(\ref{eq:Bgradient}),
\begin{equation}
    \frac{\partial B}{\partial K_{ab}}
    =\frac{B}{2}\Big[M_{ab}-\big(c_0M^{\psi}+c_1\bar{M}+c_2S\big)_{ab}\Big],
    \label{sm:eq:Bgrad}
\end{equation}
which completes the proof of Theorem~\ref{thm:bound-iteration}. We have
checked Eq.~\eqref{sm:eq:Bgrad} against central finite differences of
$B(K)$ in a Jordan--Wigner representation with $N=3$ modes, over random real
antisymmetric $K$ and Haar-random targets, with a maximal discrepancy of
$10^{-9}$.

\subsection{Consistency checks}

\paragraph{Gaussian limit.} If $\ket{\psi}$ is an eigenstate of $\Gamma$
then $V=0$, and $S_{ab}=\bra{\psi}(-\Gamma)X_{ab}\ket{\psi}=\mu
M^{\psi}_{ab}$. Hence $\alpha=e^{\mu}$, $c_1=0$,
$c_0=(e^{\mu}-2\mu h)/e^{\mu}$ and $c_2=2h/e^{\mu}$, so that
\begin{equation}
    c_0M^{\psi}+c_2S
    =\frac{e^{\mu}-2\mu h+2\mu h}{e^{\mu}}M^{\psi}=M^{\psi}.
\end{equation}
The Gaussian projection is a fixed point, as it must be, and the bound is
saturated. This check fails under the opposite sign convention for $\mu$ and
is therefore the cheapest available test of an implementation.

\paragraph{Removable singularity.} As $\delta\to0$ all three of $h$,
$\partial_\lambda h$ and $\partial_\mu h$ remain finite,
Eq.~\eqref{sm:eq:removable}, so the coefficients in Theorem \ref{thm:bound-iteration} are
regular there.

\subsection{Practical evaluation}

Every ingredient of Eq.~\eqref{sm:eq:Bgrad} is available at
$\mathcal{O}(N^{3})$ cost from the canonical frame of $K$, together with
target data measured once. From Eq.~\eqref{sm:eq:canonical},
$\mathcal{Z}_K=\prod_k2\cosh(\beta_k/2)$ and
$\lambda_{\max}=\tfrac12\sum_k|\beta_k|$. The top eigenvector of $-\Gamma$
is the Bogoliubov configuration with $n_k=0$ for $\beta_k>0$ and $n_k=1$ for
$\beta_k<0$, which is itself a \emph{pure Gaussian state}; its covariance
matrix therefore follows from the canonical frame alone,
\begin{equation}
    \bar{M}=O\Big(\bigoplus_k(-\operatorname{sgn}\beta_k)\,\varepsilon\Big)O^{\mathsf T},
    \label{sm:eq:Mbar}
\end{equation}
with $O$ and $\beta_k$ as in Eq.~\eqref{sm:eq:youla}. No perturbation theory
and no many-body diagonalisation are needed. We have verified
Eq.~\eqref{sm:eq:Mbar} against the exact top eigenvector of $-\Gamma$ to
$10^{-15}$.

The target enters only through fixed tensors contracted with $K$,
\begin{equation}
\begin{aligned}
    \mu&=-\frac{1}{2}\sum_{c<d}K_{cd}\,M^{\psi}_{cd},\\
    V&=\frac{1}{4}\sum_{a<b}\sum_{c<d}K_{ab}K_{cd}
       \Big[\bra{\psi}X_{ab}X_{cd}\ket{\psi}-M^{\psi}_{ab}M^{\psi}_{cd}\Big],\\
    S_{ab}&=-\frac{1}{4}\sum_{c<d}K_{cd}\,
       \bra{\psi}\{X_{cd},X_{ab}\}\ket{\psi},
\end{aligned}
    \label{sm:eq:contractions}
\end{equation}
The two- and four-point Majorana tensors $M^{\psi}$ and
$\bra{\psi}X_{ab}X_{cd}\ket{\psi}$ are therefore measured \emph{once}; every
iteration re-contracts the same data with an updated $K$ and is classical
$\bar{M}$ costs nothing either. The spectral-edge eigenvector is itself a pure Gaussian state, fixed by the canonical form of $K$ alone through $\bar{M}=O\big(\bigoplus_k( \operatorname{sgn}\beta_k)\varepsilon\big)O^{\mathsf T}$.  No repeated access to the quantum state is required at any point, and in particular no tomography.

\subsection{Evaluation of $\mathcal{D}^{\mathrm{L}}$ and its dependence on the regulator}
\label{sec:sm-epsilon}

Theorem~\ref{thm:lower-bound} holds for every $\epsilon\in(0,1)$, so any admissible choice
already yields a valid bound, and Corollary~\ref{cor:eps} returns the best of them. We
describe here how the curves of Fig.~\ref{fig:faf_vs_d} were obtained, and show that the
result does not hinge on that choice.

For each $U$ and each $\epsilon$ on the grid
$\{10^{-5},3\!\times\!10^{-5},10^{-4},3\!\times\!10^{-4},10^{-3},2\!\times\!10^{-3},
3\!\times\!10^{-3},5\!\times\!10^{-3},10^{-2},2\!\times\!10^{-2},3\!\times\!10^{-2},
5\!\times\!10^{-2}\}$, we maximise $B$ over $\mathcal{B}_{\epsilon}$ using the update rule
of Eq.~\eqref{eq:bound-selfconsistency}. The iteration is seeded with the Gaussian
projection $\sigma^{(0)}$, projected into the box, together with seven further starts
obtained by adding symmetric Gaussian noise of amplitude $10^{-2}$ to the seed; the
perturbation also breaks the symmetries that the update would otherwise preserve. Each
start runs for at most $40$ steps, or until the correlation matrices move by less than
$10^{-12}$ in Frobenius norm, and the constraint $|m_{k}|\le1-\epsilon$ is imposed after
every update by clipping the eigenvalues of the correlation matrix to
$[\epsilon/2,\,1-\epsilon/2]$. Because Eq.~\eqref{eq:DL} requires an \emph{upper} bound on
$\sup_{\mathcal{B}_{\epsilon}}B$, we retain the largest $B$ encountered at any admissible
iterate of any start rather than the converged value; underestimating the supremum would
inflate the bound. We then report $\mathcal{D}^{\mathrm{L}}_{\epsilon}
=1-(1-\epsilon/2)^{-N}\max B$ and
$\mathcal{D}^{\mathrm{L}}=\max_{\epsilon}\mathcal{D}^{\mathrm{L}}_{\epsilon}$.
\begin{figure}
    \centering
    \includegraphics[width=0.5\linewidth]{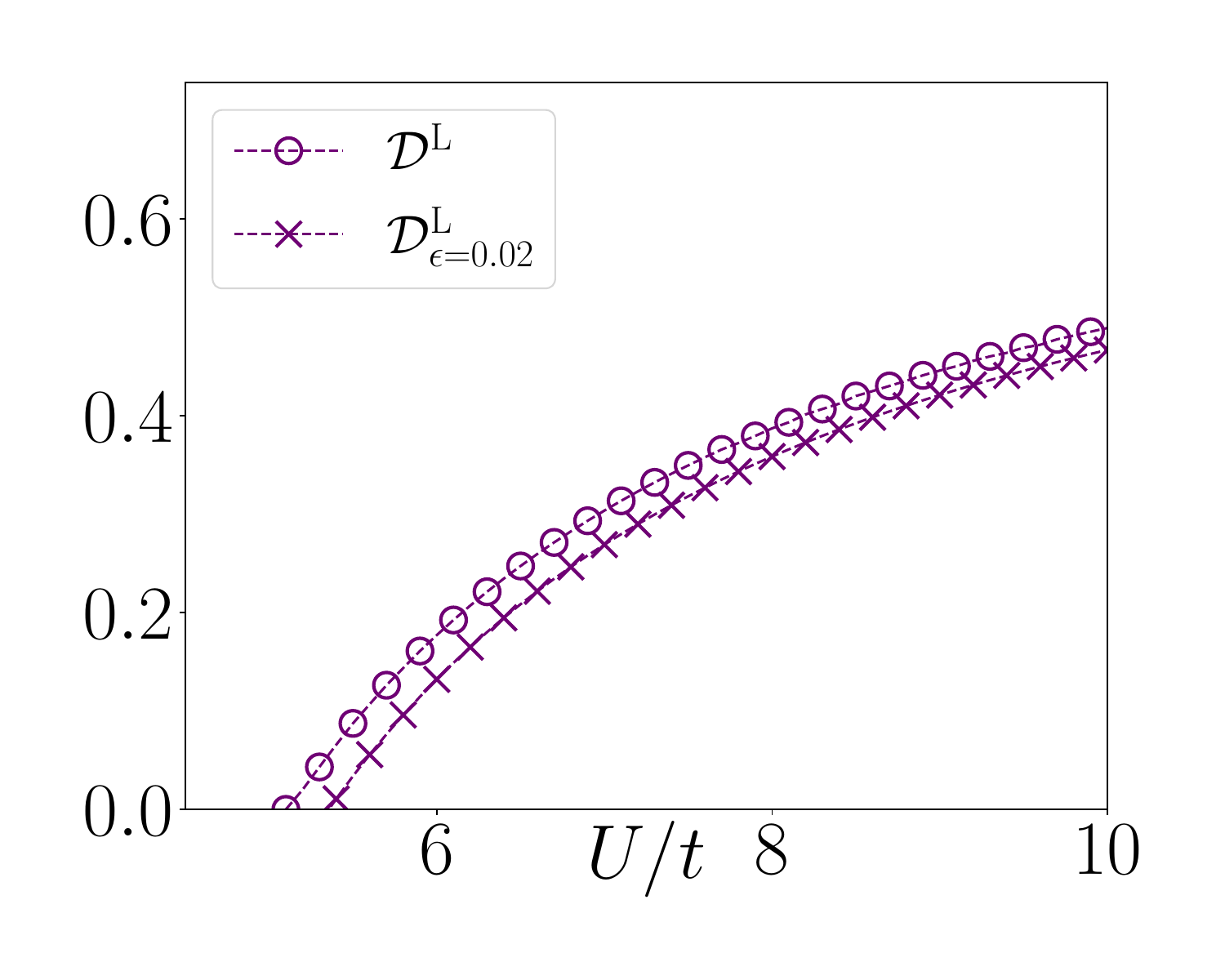}
    \caption{Dependence of the lower bound on the regulator, for the Fermi--Hubbard ground
state at $N=8$ sites and half filling. Circles show
$\mathcal{D}^{\mathrm{L}}=\sup_{\epsilon}\mathcal{D}^{\mathrm{L}}_{\epsilon}$, maximised
over the grid of $\epsilon$ given in Sec.~\ref{sec:sm-epsilon}; crosses show the bound at
the single fixed value $\epsilon=0.02$, obtained from one run of the iteration and
rescaled by the prefactor $(1-\epsilon/2)^{-N}$ of Eq.~\eqref{eq:DL}. The two lie close to
one another wherever the witness is non-trivial and converge as $U/t$ grows, so that the
choice of regulator does not materially affect the result. Since Eq.~\eqref{eq:DL} holds
for every $\epsilon$, the maximisation over $\epsilon$ buys tightness rather than
validity.}
    \label{fig:epsilon_n8}
\end{figure}
At $N=8$ the optimum sits at $\epsilon^{\star}=3\times10^{-3}$ throughout
$5.2\le U/t\le9.7$, moving to $5\times10^{-3}$ above. This is the balance implied by the
prefactor, $\mathrm{d}\log(\sup B)/\mathrm{d}\epsilon=-N/2$ at the optimum. The maximum is
shallow: at $U/t=7$ the bound takes the values $0.3013$, $0.3025$ and $0.3019$ for
$\epsilon=2,3,5\times10^{-3}$, so the reported value is stationary in $\epsilon$ rather
than tuned to it.

Figure~\ref{fig:epsilon_n8} makes the same point directly. The crosses there show the
bound at the single fixed value $\epsilon=0.02$, obtained from one run of the iteration
seeded by the Gaussian projection and rescaled by the prefactor $(1-\epsilon/2)^{-N}$ of
Eq.~\eqref{eq:DL}; it is an estimate of $\mathcal{D}^{\mathrm{L}}_{\epsilon=0.02}$ rather
than a value certified against a search over $\mathcal{B}_{\epsilon}$, but it suffices
here. The two curves lie close to one another throughout the range in which the witness is
non-trivial, and converge as $U/t$ grows. The regulator is therefore not a fitted
parameter: any $\epsilon$ in the useful window returns essentially the same bound.

\section{Update rule under particle conservation}
When the target and trial states conserve particle number, the anomalous part of $K$ vanishes. The effective Hamiltonian becomes $\Gamma=\sum_{jk}A_{jk}c^\dagger_jc_k$ with $A$ Hermitian, the covariance matrix is fixed by $C_{jk}=\bra{\psi}c^\dagger_jc_k\ket{\psi}$ through $M=(2C-\mathbb{I})\otimes\epsilon$ for real $C$. Here, $C =  (\mathbb{I}+e^{A})^{-1}$~\cite{Peschel2003Mar},
and Eq.~\eqref{eq:update} becomes the Hermitian part of
\begin{equation}
   C^{(r+1)}_{jk}
   =\frac{\bra{\psi}\,c^\dagger_jc_k\;e^{-\Gamma^{(r)}}\,\ket{\psi}}
         {\bra{\psi}\,e^{-\Gamma^{(r)}}\,\ket{\psi}}\, .
   \label{eq:iteration-dirac}
\end{equation}
Hermiticity of $C$ is imposed in any implementation, so the two formulations coincide in practice, and the Majorana form is needed only when the closest Gaussian state carries pairing.

\section{Benchmarking with Simulated Annealing}
In this section, we outline the simulated annealing protocol to obtain the closest Gaussian state to a given target state via global minimization. Simulated annealing is a non-deterministic (probabilistic) optimization algorithm that occasionally accepts bad solutions to escape local minima. Under ideal conditions, simulated annealing always converges to the global optimal solution.
\begin{figure}
    \centering
    \includegraphics[width=0.8\linewidth]{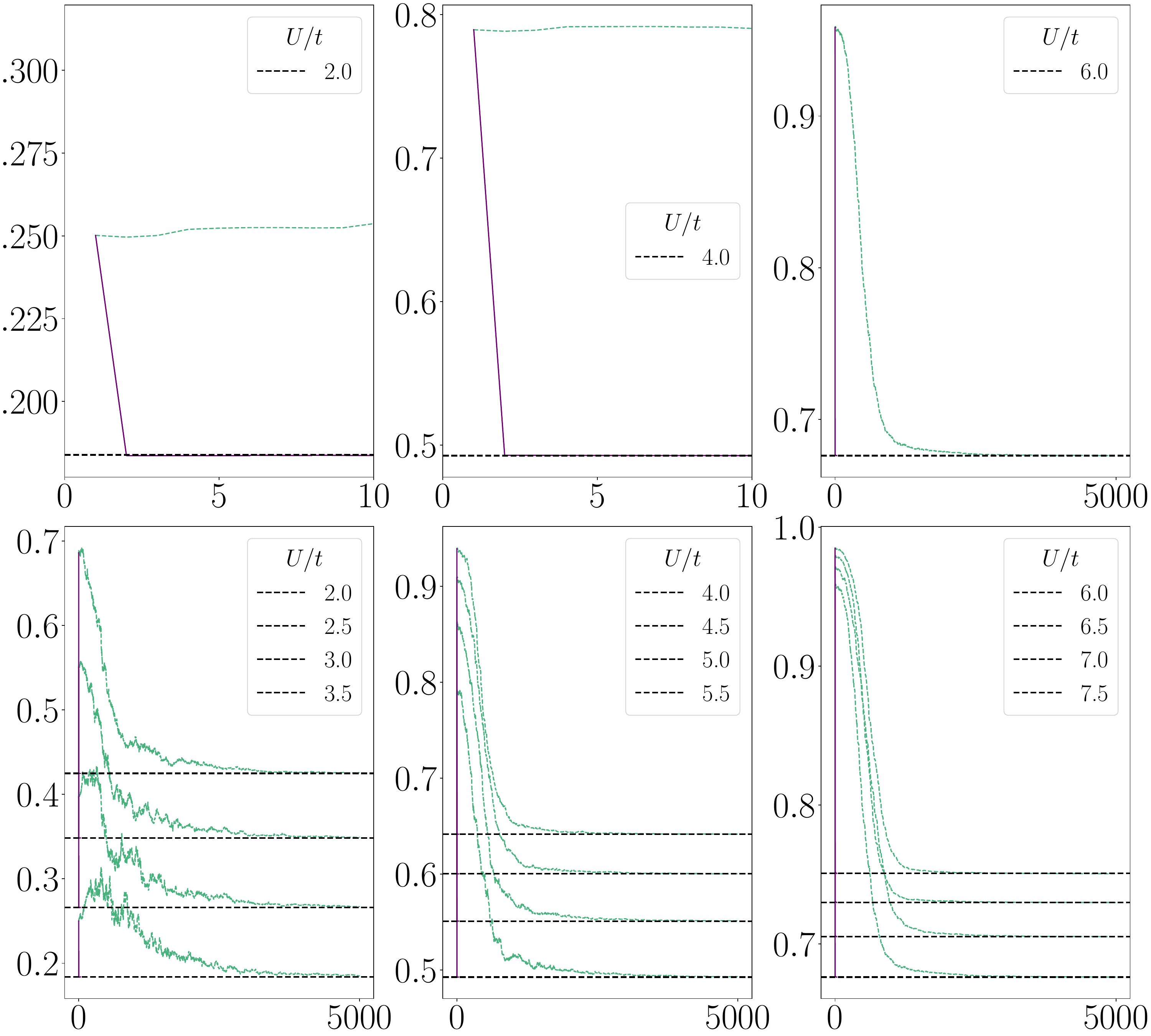}
    \caption{Comparison of the iterative algorithm with simulated annealing for the ground state of the Fermi--Hubbard model for several values of $U/t$ ($N=10$). Both methods start from the Gaussian projection $\sigma^{(0)}$, whose infidelity with the target is $1-\mathcal{F}\approx0.41$, and are compared as a function of the number of fidelity evaluations, i.e., applications of $e^{-\Gamma}$ to the target state. The iterative algorithm (solid purple line) converges within a single step to $1-\mathcal{F}\approx0.27$ (dashed black line). Simulated annealing (dashed green line; initial temperature $T_0=0.05$, geometric cooling factor $0.995$, step size $0.25$) initially accepts moves that increase the infidelity, and after $600$ evaluations reaches $1-\mathcal{F}\approx0.30$, still above the value the iteration attains after one.}
    \label{fig:SAvsit}
\end{figure}
To implement this, we first write the generic fermionic quadratic Hamiltonian in the following form:
\begin{eqnarray}
 \hat{H}&=&\sum_{i, j=1}^{N}\left( A_{ij}c^{\dagger}_{i}c_j -A^{*}_{ij} c_{i}c^{\dagger}_{j} +B_{ij}c_{i}c_{j}-B^{*}_{ij}c^{\dagger}_{i}c^{\dagger}_{j} \right)\nonumber\\
   &=& \Vec{\bm{c}}^{\dagger} Q \Vec{\bm{c}}, \quad\quad\thickspace\text{where }\thickspace 
  Q=\begin{pmatrix} -A^{*} & B \\ -B^{*} & A \end{pmatrix}
\end{eqnarray}
with $A$ being Hermitian and $B$ skew-symmetric, both of dimension $N\times N$. Moreover, $\Vec{c}$ is a vector consisting of all the creation and annihilation operators:
\begin{equation}
\Vec{\bm{c}}^{T} =
\left(
c_0^\dagger,\ c_1^\dagger,\ \cdots,\ c_{N-1}^\dagger,\ c_0,\ c_1,\ \cdots,\ c_{N-1}
\right).
\end{equation}
Then, fermionic Gaussian states $\sigma_G$ can be represented as $\sigma_G=e^{-\hat{H}}$ for some $A$ and $B$. Subsequently, we define the cost function as the distance between the initial state $\rho$ and the Gaussian state we obtain at every step of the annealing algorithm, as characterized by $A$ and $B$; 
\begin{eqnarray}
f(A, B) = 1-\mathcal{F}(\rho, \sigma_G), \quad\quad \text{ where } \sigma_G=\sigma_G(A, B)   
\end{eqnarray}
As the initial guess for the closest Gaussian state, we use the Gaussianified state of \(\rho\), obtained from its two-point correlation functions. For the numerical implementation, we use an exponential cooling schedule with an initial temperature $T_{0}=0.995$ and a cooling rate of $r=0.05$. See the following algorithm for the implementation. 

Fig.~\ref{fig:SAvsit} shows numerical results for simulated annealing, along with a comparison with those obtained using the iterative algorithm outlined in the main text.

\iffalse
The matrix $Q$ is related to the correlation matrix $\Gamma$ as follows:
\begin{equation}
   \Gamma=\dfrac{1}{1+\exp\{2Q\}}\implies Q=\dfrac{1}{2}\ln\left(\Gamma^{-1}-\mathbb{I}_{2N}\right).
\end{equation}

We make use of the above relation to get an educated guess for the closest Gaussian state for a given density matrix. In particular, given a density matrix $\rho$, one can evaluate the corresponding correlation matrix (consisting of two-point correlators) $\Gamma$. By plugging $\Gamma$ into the above relation, we obtain $Q$, which we shall use as the initial condition in the optimization algorithm.  
\fi

\begin{algorithm}[H]
\label{Alg:SA_alg}
\caption{SA algorithm to find the closest Gaussian state}
\begin{algorithmic}[1]
\State \textbf{Input:} Evaluation function \( f: \mathcal{X} \rightarrow \mathbb{R} \), initial guess \( (A_0, B_0) \in \mathcal{X} \), initial temperature \( T_0 > 0 \), cooling schedule \( T_k \), maximum iterations \( N_{\text{max}} \)
\State \textbf{Output:} Approximate minimizer \( A^*, B^{*} \in \mathcal{X} \)

\State Initialize: \( (A, B) \gets (A_0, B_0) \), \( E \gets f(A_0, B_0) \), \( k \gets 0 \)

\While{$k < N_{\text{max}}$}
    \State Generate a candidate $ (A', B') $ in the neighborhood of $(A, B)$
    \State Compute energy \( E' \gets f(A', B') \)
    \If{ \( E' < E \) }
        \State Accept: \( (A, B) \gets (A', B') \), \( E \gets E' \)
    \Else
        \State Accept with probability \( p = \exp\left( -\frac{E' - E}{T_k} \right) \)
        \If{Random uniform variable \( u \leq p \)}
            \State \( (A, B) \gets (A', B') \), \( E \gets E' \)
        \EndIf
    \EndIf
    \State Update temperature: \( T_{k+1} \gets \text{CoolingSchedule}(T_k, k) \)
    \State \( k \gets k + 1 \)
\EndWhile

\State \Return \( (A^*, B^*) \gets (A, B) \)
\end{algorithmic}
\end{algorithm}

\section{Bounds without optimisation}
\label{app:fixed-reference}
\begin{figure}
    \includegraphics[width=0.6\linewidth]{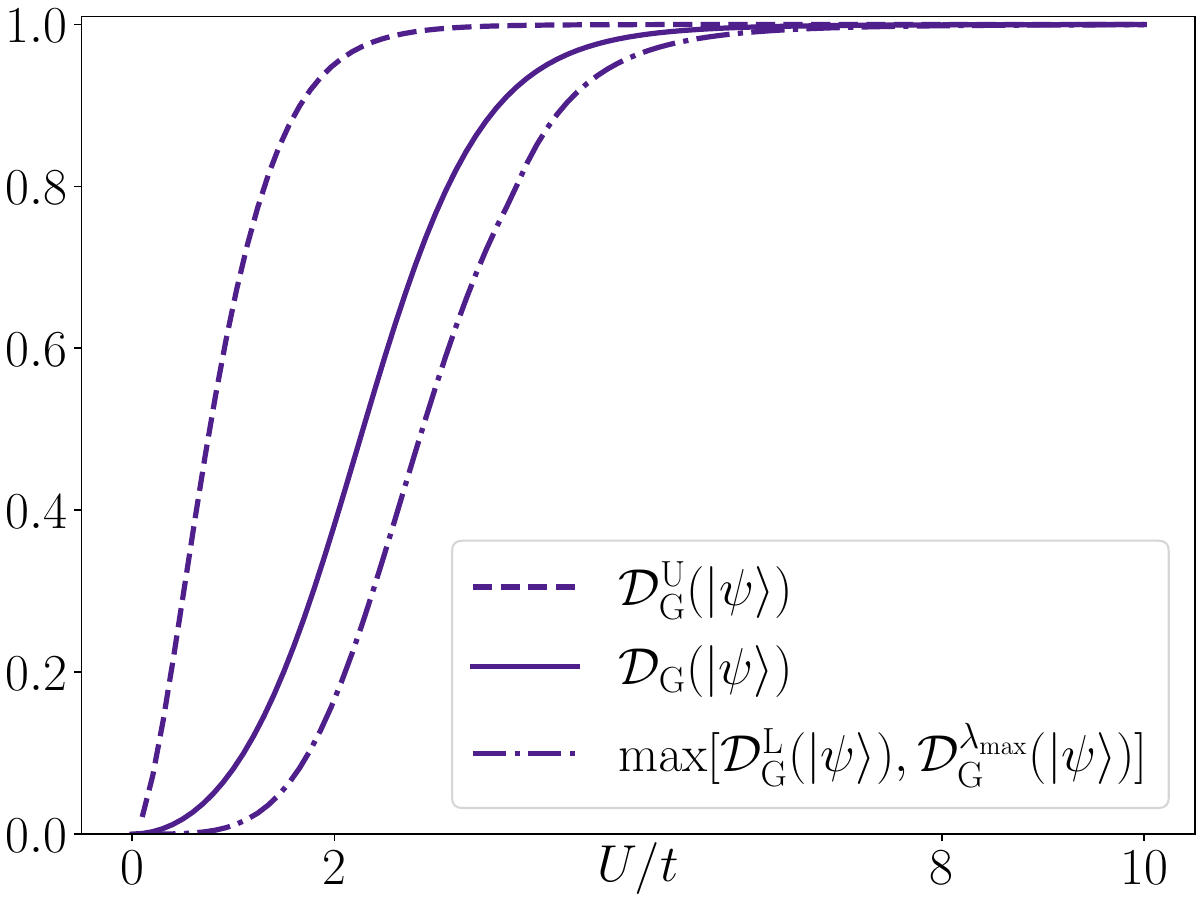}
    \caption{Bounds on the non-Gaussianity of the Fermi--Hubbard ground state obtained
    without optimisation, at half filling and $N=12$ sites. All quantities are
    evaluated at the fixed Gaussian projection $\sigma^{(0)}$, whose covariance matrix
    matches that of the target. Shown are the distance to that reference,
    $\mathcal{D}_{\mathrm{G}}$(solid), together with the bracket it satisfies:
    the upper bound $\mathcal{D}^{\mathrm{U}}_{\mathrm{G}}$(dashed)  (see Eq.~\eqref{eq:fixed_upper_bound}), and the lower estimates
    $\mathcal{D}^{\mathrm{L}}_{\mathrm{G}}$ and
    $\mathcal{D}^{\lambda_{\max}}_{\mathrm{G}}$(dash-dotted) of
    Eqs.~\eqref{eq:fixed_lower_bound} and~\eqref{eq:lammax} respectively. The bracket tightens as the target
    departs from the Gaussian manifold. Note that
    $\mathcal{D}^{\mathrm{L}}_{\mathrm{G}}$ bounds $\mathcal{D}_{\mathrm{G}}$ and not
    $\mathcal{D}$; the lower bound on the measure itself is
    $\mathcal{D}^{\mathrm{L}}$ of Theorem~\ref{thm:lower-bound}, which requires the
    maximisation these estimates avoid.}
    \label{fig:gauss_fid_bounds}
\end{figure}
The bound of Theorem~\ref{thm:lower-bound} calls for a maximisation of $B$ over
$\mathcal{B}_{\epsilon}$. Simple estimates are available that require no optimisation at
all, and are therefore certified as computed. They follow from evaluating
Eq.~\eqref{eq:jensen} at a single fixed Gaussian reference, for which the natural choice is
the Gaussian projection $\sigma^{(0)}$, the Gaussian state whose covariance matrix matches
that of the target, and the seed of the iteration of Eq.~\eqref{eq:update}. Writing
$K^{(0)}$ for its coefficient matrix and evaluating $\mu$, $\operatorname{Var}_\psi(\Gamma)$,
$\lambda_{\min/\max}$ and $\mathcal{Z}_{K}$ there throughout,
\begin{align}
    \mathcal{D}^{\mathrm{L}}_{\mathrm{G}}&=1-B\big(K^{(0)}\big),
    \text{ and } \label{eq:fixed_lower_bound} \\
    \mathcal{D}^{\mathrm{U}}_{\mathrm{G}}
      &=1-\frac{e^{\mu}+h(\lambda_{\min};\mu)\operatorname{Var}_\psi(\Gamma)}
              {\mathcal{Z}_{K}}\bigg|_{K=K^{(0)}} 
    \label{eq:fixed_upper_bound}
\end{align}
the two sides of Eq.~\eqref{eq:jensen} taken at the same reference. A third estimate uses
only the spectrum of $\sigma^{(0)}$: since
$\bra{\psi}e^{-\Gamma}\ket{\psi}\le e^{\lambda_{\max}}$ for any normalised $\ket{\psi}$,
\begin{equation}
    \mathcal{D}^{\lambda_{\max}}_{\mathrm{G}}
      =1-\frac{e^{\lambda_{\max}}}{\mathcal{Z}_{K}} .
    \label{eq:lammax}
\end{equation}
Equation~\eqref{eq:lammax} admits no minimisation. Driving $\lambda_{\max}$ down across the
manifold sends it to trivial values, so it serves as a bound at a fixed reference state
rather than as a variational quantity.

These estimates bracket the distance to the Gaussian projection,
$\mathcal{D}_{\mathrm{G}}(\ket{\psi})=1-\bra{\psi}\sigma^{(0)}\ket{\psi}$, through two
chains of inequalities,
\begin{align}
    \mathcal{D}^{\mathrm{L}}
      &\;\le\;\mathcal{D}(\ket{\psi})
      \;\le\;\mathcal{D}_{\mathrm{G}}(\ket{\psi})
      \;\le\;\mathcal{D}^{\mathrm{U}}_{\mathrm{G}},
    \label{eq:chain1}\\
    \max\big[\mathcal{D}^{\mathrm{L}}_{\mathrm{G}},\,
             \mathcal{D}^{\lambda_{\max}}_{\mathrm{G}}\big]
      &\;\le\;\mathcal{D}_{\mathrm{G}}(\ket{\psi})
      \;\le\;\mathcal{D}^{\mathrm{U}}_{\mathrm{G}} .
    \label{eq:chain2}
\end{align}
The first holds because $\mathcal{D}$ is an infimum, so that any single Gaussian state
bounds it from above, while Theorem~\ref{thm:lower-bound} bounds it from below; the second
follows by applying Eq.~\eqref{eq:jensen} at $K^{(0)}$.
Figure~\ref{fig:gauss_fid_bounds} shows the bounds given in Eq.~\eqref{eq:chain2}, which require no optimization on the Gaussian state. The lower bound $\max[\mathcal{D}^{\mathrm{L}}_{\mathrm{G}}, \mathcal{D}^{\lambda_{\max}}_{\mathrm{G}}]$ tracks $\mathcal{D}_{\mathrm{G}}$ quite closely over the entire range of interactions, and it remains finite for all $U/t$. As the two constituents are informative in complementary regimes, the spectral bound at weak coupling and the variance bound at strong coupling, their maximum is tight at both limits. Importantly, these curves bound the distance to the Gaussian projection of the target, not to the closest Gaussian state, making them not a strict witness but enabling them to behave better at weak coupling than $\mathcal{D}^{\mathrm{L}}$ does.

Two features are worth separating. $\mathcal{D}_{\mathrm{G}}$ itself is the tight upper bound on $\mathcal{D}$, and it is directly measurable: matchgate shadows return $\bra{\psi}\sigma_{K}\ket{\psi}$ for any Gaussian $\sigma_{K}$ from the same snapshots~\cite{Wan2023}, so no iteration is needed to certify how far a prepared state sits from the free-fermion manifold. $\mathcal{D}^{\mathrm{U}}_{\mathrm{G}}$ is looser, but asks only for the first two moments of $\Gamma$. Conversely, $\mathcal{D}^{\mathrm{L}}_{\mathrm{G}}$ bounds $\mathcal{D}_{\mathrm{G}}$ from below and \emph{not} $\mathcal{D}$: no general ordering holds between the two, since the relaxation that produces $B$ and the restriction to a single reference act in opposite directions. Only $\mathcal{D}^{\mathrm{L}}$ of Theorem~\ref{thm:lower-bound}, which maximises $B$ over $\mathcal{B}_{\epsilon}$, lower-bounds $\mathcal{D}$ itself.

\end{document}